\documentclass[11pt]{iopart}
\usepackage{graphicx}

\begin{document}

\title[Nodal points and the transition from ordered to chaotic Bohmian trajectories]
{Nodal points and the transition from ordered to chaotic Bohmian trajectories}
\author{C Efthymiopoulos, C Kalapotharakos and G Contopoulos}
\address{
      Research Center for Astronomy and Applied Mathematics, Academy of Athens,
      Soranou Efesiou 4, GR-115 27 Athens, Greece
           }
\ead{cefthim@academyofathens.gr, ckalapot@phys.uoa.gr, gcontop@academyofathens.gr}

\begin{abstract}

We explore the transition from order to chaos for the Bohmian trajectories
of a simple quantum system corresponding to the superposition
of three stationary states in a 2D harmonic well with incommensurable
frequencies. We study in particular the role of nodal points in the
transition to chaos. Our main findings are: a) A proof of the existence
of bounded domains in configuration space which are devoid of nodal
points, b) An analytical construction of formal series representing
regular orbits in the central domain as well as a numerical investigation
of its limits of applicability. c) A detailed exploration of the phase-space
structure near the nodal point. In this exploration we use an adiabatic
approximation and we draw the flow chart in a moving frame of reference
centered at the nodal point. We demonstrate the existence of a saddle point
(called X-point) in the vicinity of the nodal point which plays a key role
in the manifestation of exponential sensitivity of the orbits. One of the
invariant manifolds of the X-point  continues as a spiral terminating at
the nodal point. We find cases of Hopf bifurcation at the nodal point and
explore the associated phase space structure of the nodal point - X-point
complex. We finally demonstrate the mechanism by which this complex
generates chaos. Numerical examples of this mechanism are given for
particular chaotic orbits, and a comparison is made with previous
related works in the literature.
\end{abstract}

\pacs{05.45.Mt -- 03.65.Ta}
\maketitle
%
\section{Introduction}

The formulation of quantum mechanics based on Bohm's trajectories
(de Broglie 1926, Bohm 1952a,b) has attracted considerable interest in
recent years because it offers a powerful tool to visualize quantum
processes in terms of quantum orbits. According to the Bohmian
interpretation (see Bohm and Hiley 1993 and Holland 1993 for reviews)
the particles are guided by the Schr\"{o}dinger field via deterministic
equations of motion. In this approach the moving particles,
together with the Schr\"{o}dinger field, form the basic ingredients of
objective realilty at the quantum level. On the other hand, in orthodox
quantum mechanics the orbits do not refer to deterministic particles'
motions, but they represent the streamlines of the probability current
${\mathbf j}=(\psi^*\nabla\psi-\psi\nabla\psi^*)/2i$ (we set $\hbar=1$).
Thus they represent a `Lagrangian' (Holland 2005) or `hydrodynamical'
(Madelung 1926) description of the quantum probability flow, while the
Schr\"{o}dinger equation yields the Eulerian description of the same
flow. At any rate, the descriptive power of the Bohmian approach is
independent of its ontological interpretation. In fact, applications of
Bohm's trajectories in the literature have so far successfully addressed
such basic quantum processes as the two-slit experiment (Phillipidis 1979),
the spin measurement via Stern-Gerlach devices (Dewdney et al. 1986),
tunneling through potential barriers (Hirschfelder et al. 1974, Skodje
et al. 1989, Lopreore and Wyatt 1999), ballistic electron transport
(Beenakker and van Houten 1991), superfluidity (Feynman Lectures,
Feynman et al. 1963), etc. (see Wyatt 2005 for a review of applications
of quantum dynamics with trajectories).

In the present paper we focus on one particular aspect of Bohm's
theory that refers to the distinction of the Bohmian trajectories
into {\it regular and chaotic}. A number of studies in the literature
(e.g. D\"{u}rr et al. 1992, Faisal and Schwengelbeck 1995, Parmenter and
Valentine 1995, de Polavieja 1996, Dewdney and Malik 1996, Iacomelli and
Pettini 1996, Frisk 1997, Konkel and Makowski 1998, Wu and Sprung 1999,
Makowski et al. 2000, Cushing 2000, Falsaperla and Fonte 2003, de Sales
and Florencio 2003, Wisniacki and Pujals 2005, Valentini and Westman 2005,
Efthymiopoulos and Contopoulos 2006) have so far converged to the conclusion
that generic quantum systems of more than one degrees of freedom are
characterized by the coexistence, in the configuration space, of both
regular and chaotic orbits. Some established results regarding this
distinction are:

a) The regular or chaotic character of the quantum mechanical orbits does not
necessarily correlate with the character of the classical orbits of the same
system, i.e.,  there are examples of systems with classically regular and
quantum mechanically chaotic orbits, or vice versa (see Efthymiopoulos and
Contopoulos 2006 for a review).

b) The emergence of chaos is associated with the existence of `nodal points'
in the configuration space,  i.e., points of the configuration space at which
the Schr\"{o}dinger field becomes null. In some particular examples it has
been possible to identify a mechanism by which the approach of an orbit
near a nodal point introduces chaos (Makowski et al. 2000, Wisniaski and
Pujals 2005). However the general problem of the mechanism of transition
from order to chaos for the Bohmian orbits is still largely unexplored.

c) In some cases it has been shown that the extent of chaos is related to the
number and spatial distribution of the nodal points in the configuration space
(Frisk 1997, Wisniacki et al. 2006). However, in other cases we find the
manifestation of strong chaos even if only one nodal point is present.
This problem is important because it has been shown that, when the degree
of chaos is large, it is possible to obtain an asymptotic convergence of
the distribution $p$ of an ensemble of Bohmian trajectories to Born's rule
$p=|\psi|^2$, via a Bohm-Vigier (1954) stochastic mechanism (Valentini and
Westman 2005, Efthymiopoulos and Contopoulos 2006), without having to
postulate this rule.

Our purpose in the present paper is to study how the transition from regular
to chaotic motion takes place by considering the orbits of a very simple quantum
system, namely the superposition of three eigenstates in a Hamiltonian system
of two independent harmonic oscillators. Parmenter and Valentine (1995,1996)
demonstrated that when the two oscillator frequencies are incommensurable,
the configuration space is filled by both regular and chaotic orbits (the
initial claim that all the orbits were chaotic (Parmenter and Valentine 1995)
was corrected by Parmenter and Valentine (1996), and by Efthymiopoulos and
Contopoulos (2006)). The same coexistence was found by Wisniacki and Pujals
(2005) when the frequencies are commensurable but the ampitudes of the
superposed eigenfunctions have a complex ratio, and by Konkel and Makowski (1998)
in the case of a particle in a box with infinite walls. In the above cases
there is only one nodal point which influences the orbits and introduces chaos.
Our aim below is, then, to study the transition from order
to chaos from two different points of view: a) topologically, i.e. we seek to
distinguigh which domains of initial conditions lead to regular or chaotic motion
and what theoretical criteria can be devised in order to separate these domains,
and b) dynamically, i.e., we seek to identify the dynamical mechanism behind the
transition from order to chaos.

The following is an outline of the paper and of the main results:

a) Section 2 contains a proof of the existence of domains in configuration
space where nodal points cannot appear. In our considered example the size of
these domains depends on the relative amplitudes of the three eigenfunctions
and specific quantitative estimates of this dependence are given, which are
compared to numerical results.

b) In their domain of analyticity (i.e. far from nodal points), the equations
of motion admit solutions expandable in series of a properly defined
small parameter (section 3). The series' terms can be determined by an
iterative algorithm. The resulting solutions define theoretical orbits
which are, by definition, regular. The theoretical orbits explain all the
basic characteristics of the regular orbits as found by numerical integration.
In particular they explain the frequencies, the form, the limits and the
inner deflections of regular orbits.

c) Section 4 passes to the other limit, of motion, close to the nodal points.
In order to unravel the mechanism by which the orbits approaching the nodal point
become chaotic, the key point is to take into account the motion of the nodal point
itself by passing to a description of the orbits in a {\it moving}
frame of reference centered at the (moving) nodal point. The main characteristics
of the orbits in this frame are found by expansions of the equations of motion in
terms of a new small parameter, i.e., the distance $R=\epsilon$ from the nodal point.
The angular frequency of motion near the nodal point is of order $O(1/\epsilon^2)$,
a fact allowing us to use an adiabatic approximation. In this approximation,
the flow lines near the nodal point are spirals terminating at the nodal point.
However, the flow further away from the nodal point is quite complicated and
it is studied in detail. In particular, we demonstrate how this flow is related
to the manifestation of chaos in the system. The main results are derived
theoretically and then substantiated by detailed numerical experiments.

d) Finally, we discuss (section 5) how do our results compare with previous works in
the literature on nodal points, and we end with the conclusions (section 6).

\section{Limits of nodal lines}

We study the quantum orbits in the Hamiltonian model of two uncoupled
oscillators:
\begin{equation}\label{ham2dharm}
H={1\over 2}(p_x^2+p_y^2) + {1\over 2}(x^2 + (c y)^2)
\end{equation}
when the guiding field is the superposition of three stationary states
(Parmenter and Valentine 1995):
\begin{equation}\label{eigenharm}
\psi(x,y,t) = e^{-{x^2+cy^2\over 2}-i{(1+c)t\over 2}}\big(
1+axe^{-it}+bc^{1/2}xye^{-i(1+c)t}\big)~~.
\end{equation}
The equations of motion (de Broglie 1926, Bohm 1952a) are
\begin{equation}\label{pilot}
(\dot{x},\dot{y})\equiv\Im\bigg({\vec{\nabla}\psi\over\psi}\bigg)
\end{equation}
where $\Im$ denotes the imaginary part of each of the components of the vector
$\vec{\nabla}\psi/\psi$, or
\begin{eqnarray}\label{eqmo}
{dx\over dt} &= &-{a\sin t + bc^{1/2}y\sin(1+c)t\over G} \\ \nonumber
{dy\over dt} &= &-{bc^{1/2}x\big(ax\sin ct + \sin(1+c)t\big)\over G}
\end{eqnarray}
with
\begin{equation}\label{gi}
G = 1 + 2ax\cos t + 2bc^{1/2}xy\cos (1+c)t
+ a^2x^2 + 2abc^{1/2}x^2y\cos ct + b^2cx^2y^2~~.
\end{equation}
The equations of motion (\ref{eqmo}) become singular whenever $G=0$.
From (\ref{gi}) we then find the equations of the nodal points:
\begin{equation}\label{nodal}
x_0 = -{\sin(1+c)t\over a\sin ct},~~~~
y_0 = -{a\sin t\over bc^{1/2}\sin (1+c)t}~~.
\end{equation}
When the frequency $c$ is a rational number, the nodal points describe
periodic motions in the configuration space $(x,y)$, along a finite
number of {\it nodal lines} (Figure 1a). However, when $c$ is irrational
there is an infinite number of nodal lines that fill open domains of the
space $(x,y)$ (Figure 1b). As shown in section 3, a theoretical approximation
of the regular orbits can be obtained in the complement of the domain of
nodal lines. We thus first provide, in this section, rigorous bounds for the
domain of nodal lines. In particular, we prove the following proposition:
{\it when the relative amplitudes $a,b$ are non-zero, the domain of nodal
lines is bounded by a set of limiting hyperbolae.}
    \begin{figure}\label{harnod}
    \centering
    \includegraphics[width=\textwidth]{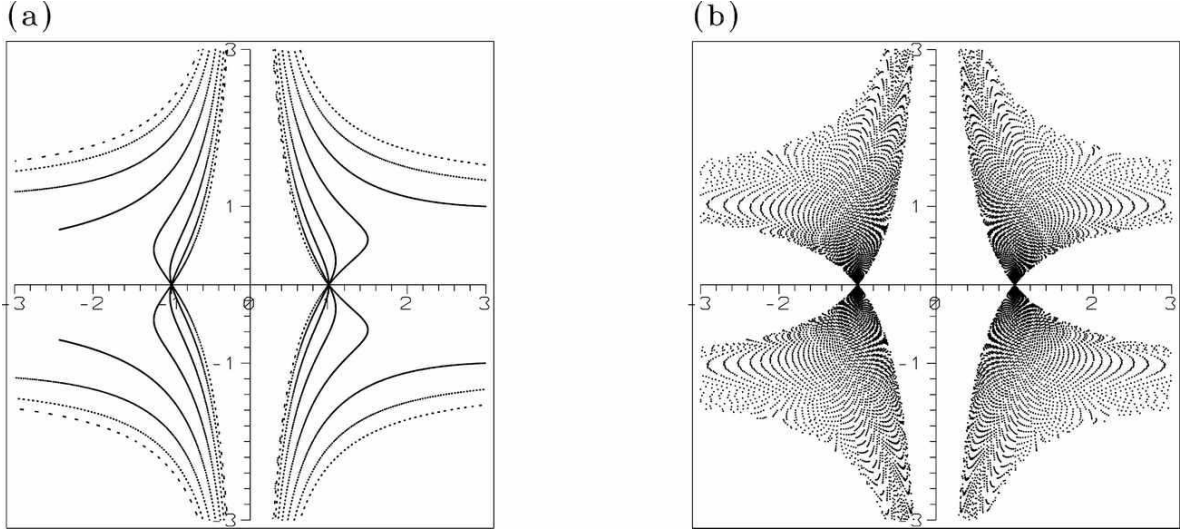}
       \caption{
Nodal lines given by Eqs.(\ref{nodal}) for $a=1$, $b=1$,
$0\leq t\leq 1000$ and (a) $c=7/10$, (b) $c=\sqrt{2}/2$.
}
    \end{figure}

    \begin{figure}\label{handlim}
    \centering
    \includegraphics[width=14cm]{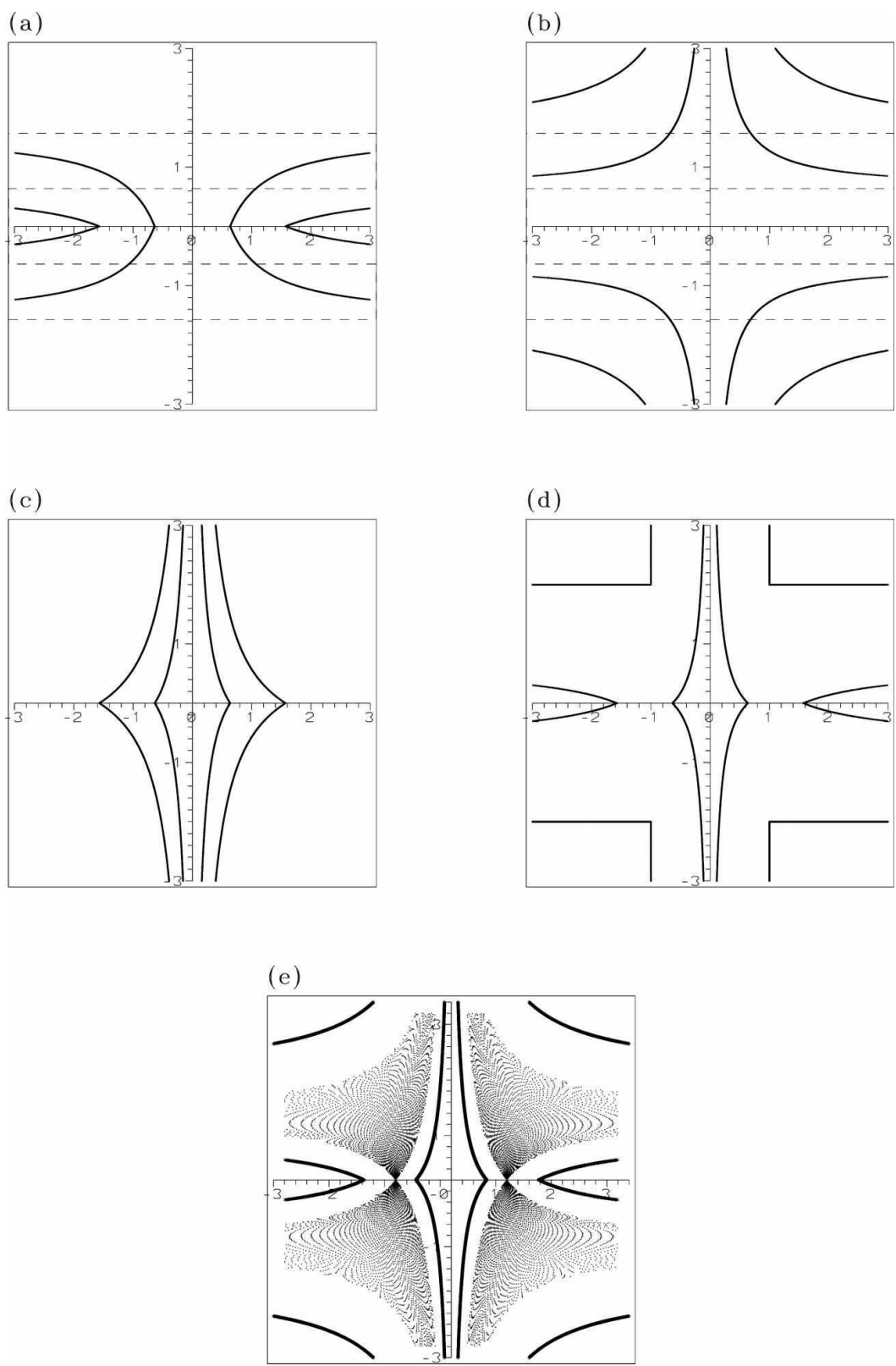}
       \caption{
The limits of nodal lines as determined by (a) Eq.(\ref{caseaun}),
(b) Eq.(\ref{casebun}), (c) Eq.(\ref{casecun}), (d) Eqs.(\ref{casedun1}) and
(\ref{casedun2}). (e) The nodal lines given by
Eq.(\ref{nodal}) for $a=1$, $b=1$, $c=\sqrt{2}/2$, $0\leq t\leq 1000$,
superposed to the less restrictive of all the limits shown in panels
(a) to (d).
}
    \end{figure}


To this end, we write $x_0=-X_0/a$ and $y_0=-aY_0/bc^{1/2}$, where
\begin{equation}\label{xynabs}
X_0 = {\sin(1+c)t\over \sin ct},~~~~
Y_0 = {\sin t\over \sin (1+c)t}~~.
\end{equation}
We make use of the following lemma:
\begin{eqnarray}\label{ubound}
&\mbox{(i)} &\forall~u\in{\cal R}~\mbox{with}~0\leq u\leq\pi/2,~~
\exists~\delta>0~\mbox{with}~{2\over\pi}\leq\delta\leq 1:~
\sin u=\delta u\nonumber\\
&\mbox{(ii)} &\forall~u\in{\cal R}~\mbox{with}~0\leq u\leq\pi,~~
\sin u\leq u~~.
\end{eqnarray}
In order to find the bounds of $(|X_0|,|Y_0|)$ we consider all the $4\times 4=16$
possible combinations of the trigonometric arguments $t,(1+c)t$ being in any
of the four quartiles of the trigonometric circle, i.e.,
\begin{eqnarray}\label{trigcomb}
(1+c)t &= &2k_1\pi\pm t_1,~\mbox{or}~(1+c)t=(2k_1+1)\pi\pm t_1\nonumber\\
t&=&2k_2\pi\pm t_2,~\mbox{or}~t=(2k_2+1)\pi\pm t_2
\end{eqnarray}
with $(k_1,k_2)\in{\cal Z}^2$ and $0\leq t_1\leq \pi/2$, $0\leq t_2\leq \pi/2$.
Taking now $ct=(1+c)t-t$ we find the following possibilities for $|\sin(ct)|$:
\begin{equation}\label{sinct}
|\sin(ct)| = |\sin(t_1\pm t_2)|~~.
\end{equation}
We then distinguish four cases, namely cases A,B with $|\sin ct|=|\sin(t_1-t_2)|$,
and cases C,D with $|\sin ct|=|\sin(t_1+t_2)|$.

{\it Case A:} $t_1>t_2$ (and $|\sin(ct)| = \sin(t_1- t_2)$).
Since $t_1-t_2\leq \pi/2$, according to the Lemma (\ref{ubound})(i)
$$
\exists~\delta_i>0,~i=1,2,3:~{2\over\pi}\leq\delta_i\leq 1~\mbox{such that}~
$$
$$
|\sin(1+c)t|=\delta_1t_1,|\sin t|=\delta_2t_2,~
|\sin ct|=\sin(t_1- t_2)=\delta_3(t_1-t_2)
$$
or
\begin{equation}\label{casea}
|X_0|={\delta_1t_1\over\delta_3(t_1-t_2)},~~
|Y_0|={\delta_2t_2\over\delta_1t_1}~~.
\end{equation}
From (\ref{casea}) then follows the inequality:
\begin{equation}\label{caseaun}
{1\over {\pi\over 2}-|Y_0|}\leq |X_0|\leq {1\over{2\over\pi}-|Y_0|}
\end{equation}
which is shown graphically in Fig.2a. When $|Y_0|=0$, $|X_0|$ is between
$2/\pi$ and $\pi/2$.

{\it Case B:} $t_2>t_1$ (and $|\sin(ct)| = \sin(t_2- t_1)$).
Working in the same way as for Case A we find the inequality
\begin{equation}\label{casebun}
{1\over {\pi\over 2}|Y_0|-1}\leq |X_0|\leq {1\over{2\over\pi}|Y_0|-1}
\end{equation}
shown graphically in Fig.2b. When $|X_0|\rightarrow\infty$, $|Y_0|$ is between
$2/\pi$ and $\pi/2$ (dashed asymptotic curves in Fig.2b).

{\it Case C:} $t_1+t_2\leq \pi/2$ (and $|\sin(ct)| = \sin(t_1+t_2)$).
In this case we find that
$$
\exists~\delta_i>0,~i=1,2,3:~{2\over\pi}\leq\delta_i\leq 1~\mbox{such that}~
$$
$$
|\sin(1+c)t|=\delta_1t_1,|\sin t|=\delta_2t_2,~
|\sin ct|=\sin(t_1+ t_2)=\delta_3(t_1+t_2)
$$
or
\begin{equation}\label{casecun}
{2\over \pi(|Y_0|+1)}\leq |X_0|\leq {\pi\over 2(|Y_0|+1)}
\end{equation}
(Fig.2c). When $|Y_0|=0$, $|X_0|$ is between $2/\pi$ and $\pi/2$.

{\it Case D:} $t_1+t_2>\pi/2$ (and $|\sin(ct)| = \sin(t_1+t_2)$).
In this case
$$
\exists~\delta_i>0,~i=1,2,3:~{2\over\pi}\leq\delta_i\leq 1~\mbox{such that}~
$$
$$
|\sin(1+c)t|=\delta_1t_1,|\sin t|=\delta_2t_2,~
|\sin ct|=\sin(t_1+ t_2)=\sin(\pi-(t_1+ t_2))=\delta_3(\pi-t_1-t_2)
$$
implying
\begin{equation}\label{casedun1}
{2\over \pi({\pi\over 2}|Y_0|+1)}\leq |X_0|\leq {1\over {2\over\pi}-|Y_0|}~~.
\end{equation}
Furthermore, if $t_1\leq\pi/4$ (case D$_1$) we have
$t_2\geq\pi/4$ and $\pi-(t_1+t_2)\geq \pi/4$. The Eqs.(\ref{xynabs}) and
(\ref{trigcomb}) imply that $|X_0|\leq 1$ and $|Y_0|\geq 1$. On the other
hand, if $t_1>\pi/4$ (case D$_2$) we have
$$
|Y_0|={\sin t\over\delta_1t_1}<{1\over\delta_1{\pi\over 4}}<2~~.
$$
The union of the two possibilities yields:
\begin{equation}\label{casedun2}
|X_0|\leq 1~~~\mbox{or}~~~|Y_0|\leq 2~~.
\end{equation}
The inequalities (\ref{casedun1}) and (\ref{casedun2}) are shown graphically
in Fig.~(2d). When $|Y_0|=0$, $|X_0|$ is between $2/\pi$ and $\pi/2$.

The less restrictive of all bounds determine the permissible domain for nodal
points. The outer limit (\ref{casebun}) of Case B is less restrictive than the
limit $|X_0|$ of Case D if $1/({2\over\pi}|Y_0|-1)<1$, i.e. if $|Y_0|<\pi$
(in the exceptional case that $|Y_0|>\pi$ the limit $|X_0|=1$ is less restrictive
than the limit of case B).
Figure 2e shows the analytical estimates for the bounds of the nodal lines compared
to the numerical determination of the nodal lines in the variables $(x_0,y_0)$ for
the parameters $a=b=1$, $c=\sqrt{2}/2$. The further restrictions of the cases A-D are
also satisfied if $t_1$ and $t_2$ take the values specified in these particular cases.
The main conclusion is that there is a central domain, with boundary defined by the
innermost arcs of limiting hyperbolae, that is never crossed by nodal points. There
is also an outer domain, at large distances from the center, which is again prohibited
to nodal points. By calculating many orbits, our numerical evidence is that the orbits
lying within these domains are regular. In particular, we now turn our attention to
the analytical determination of the regular orbits for the central domain free of
nodal points.

\section{Integrals of motion and the bounds of regular orbits}

According to the analysis of the previous section, a lower bound for the
distance of a nodal point $(x_0,y_0)$ from the origin $(0,0)$ is given by
$d_{min}=(x_{0,min}^2+y_{0,min}^2)^{1/2}$, with $(x_0=X_{0,min}/a,
y_0=aY_{0,min}/bc^{1/2})$, and $X_{0,min},Y_{0,min}$ corresponding to the closest
approach of the innermost limiting hyperbola (Eq.(\ref{casedun1})) to the origin,
i.e.
$$
{\pi^4X_{0,min}^4\over 16}+{\pi X_{0,min}\over 2}-1=0,~~
Y_{0,min}={2\over\pi}\Bigg({2\over \pi X_{0,min}}-1\Bigg)
$$
or $X_{0,min}=0.461226$, $Y_{0,min}=0.242092$.
When $a$ and $b$ are large, the distance $d_{min}$ is small and
the inner domain free of nodal points is small. On the other hand, in the
limit $a\rightarrow 0$, $b\rightarrow 0$ we have $d_{min}\rightarrow
\infty$ and the whole space is free of nodal points. This implies that
if we look for analytic solutions of the equations of motion (\ref{eqmo}),
i.e., free of singularities in the neighborhood of the origin, we may
consider $a,b$ in Eq.(\ref{gi}) as small parameters, and expand $1/G$
in Eq.(\ref{eqmo}) in a power series of these parameters. This results
in solutions
\begin{equation}\label{xyexpand}
x(t)=x_0+x_1(t) + x_2(t)+...,~~~
y(t)=y_0+y_1(t) + y_2(t)+...,~~~
\end{equation}
where the functions $x_n(t)$, $y_n(t)$ are of order $n$ in the amplitudes
$a$ or $b$. The convergence of these series, which are of the form of the
`third integral' (Contopoulos 1960), is an open problem. However, our numerical
evidence below is that the form of theoretical orbits derived by these series
fits well the form of the numerical orbits for small parameters $a,b$.

In the zeroth order approximation all the solutions are equilibria $x(t)=x_0$,
$y(t)=y_0$. This corresponds to the limit $a=b=0$, in which the guiding
$\psi-$field is a bound stationary state, and all the quantum orbits
are neutral equilibria. On the other hand, when higher order terms are taken
into account, Eqs.(\ref{xyexpand}) can be inverted, and, provided that the
series converge, they yield the integrals of motion
\begin{equation}\label{xyinte}
x-x_1(t)-x_2(t)-...=x_0,~~~
y-y_1(t)-y_2(t)-...=y_0
\end{equation}
implying that the resulting orbits are, by definition, regular.

The solutions (\ref{xyexpand}) are found recursively, i.e., order by
order, giving $x_k,y_k$ as explicit trigonometric expressions in $t$,
with two basic frequencies $\nu_1=1$ and $\nu_2=c$.

The first order equations read
\begin{equation}\label{h2eqmo1}
{dx_1\over dt} = -a\sin t - {bc^{1/2}}y_0\sin((1 + c)t),~~~
{dy_1\over dt} = -bc^{1/2}x_0\sin((1 + c)t)
\end{equation}
and they can be readily integrated yielding
\begin{eqnarray}\label{h2sol1}
x_1(t) &= &A_1 + a\cos t + bc^{1/2}y_0{\cos((1 + c)t)\over 1 + c}
\nonumber \\
y_1(t) &= &B_1 + bc^{1/2}x_0{\cos((1 + c)t)\over 1 + c}
\end{eqnarray}
where $A_1,B_1$ are integration constants. Since we wish to absorb the whole
dependence of the solution on the initial conditions in the $x_0,y_0$ part
of the solution, we select the values of $A_1,B_1$ such that $x_1(0)=
y_1(0)=0$.

In a similar way we treat the second order equations, finding the solutions
\begin{eqnarray}\label{h2sol2x}
x_2(t) &= &-{a^2x_0\over 2}\cos(2t)-{b^2cx_0\over (1+c)^2}
\cos((1+c)t) \nonumber\\
&- &{b^2cx_0\over 2(1+c)}\bigg(y_0^2-{1\over 2(1+c)}\bigg)\cos(2(1+c)t)\\
&-&{2abc^{1/2}x_0y_0\over 2+c}\cos((2+c)t) + A_2   \nonumber
\end{eqnarray}
and
\begin{eqnarray}\label{h2sol2y}
y_2(t) &= &-\bigg({abc^{1/2}\over 1+c}+{b^2cy_0\over (1+c)^2}\bigg)
\cos((1+c)t)\nonumber\\
&-&{b^2cy_0\over 2(1+c)}\bigg(x_0^2-{1\over 2(1+c)}\bigg)
\cos(2(1+c)t)
\\
&+ &{ab\over 2c^{1/2}}\cos(ct)
-{abc^{1/2}\over 2+c}\bigg(x_0^2-{1\over 2}\bigg)
\cos((2+c)t) + B_2        \nonumber
\end{eqnarray}
respectively. The integration constants $A_2,B_2$ are also given values such
that $x_2(0)=y_2(0)=0$.

We can prove the {\it consistency} of the above construction, i.e., that no
secular terms can appear in the above recursive scheme. The proof follows by
induction: If the solutions $x_i(t),y_i(t)$, $i=1,...,n$ contain only cosine
terms (of the form $\cos((m_1+m_2c)t)$ with $m_1,m_2$ integer, then, the
expansion
$$
{1\over G} = 1+\sum_{k=1}^{n}(-1)^k\Bigg[2ax^{(n)}\cos t
+ 2bc^{1/2}x^{(n)}y^{(n)}\cos (1+c)t
$$
$$
+ a^2(x^{(n)})^2 + 2abc^{1/2}(x^{(n)})^2y^{(n)}\cos ct
+ b^2c(x^{(n)})^2(y^{(n)})^2\Bigg]^k +\ldots
$$
with
$$
x^{(n)}(t)=x_0(t)+x_1(t)+\ldots +x_n(t),~~y^{(n)}(t)=y_0(t)+y_1(t)+\ldots +y_n(t)
$$
contains only powers of cosine terms, yielding again only cosine terms since
$\cos w_1 \cos w_2 =$ $(\cos(w_1+w_2)+\cos(w_1-w_2))/2$. Thus, the equations
of motion in order $n+1$ yield only sine terms, since $1/G$ in Eqs.(\ref{eqmo})
is multiplied only by sine terms and $\sin w_1 \cos w_2 =$ $(\sin(w_1+w_2)
+\sin(w_1-w_2))/2$. If for some terms we have $w_1=(m_1+m_2c)t$,
$w_2=(m_1'+m_2'c)t$, the sine terms produced in the equations of motion
are $\sin(m_1+m_1'+m_2c+m_2'c)t$, or $\sin(m_1-m_1'+m_2c-m_2'c)t$. If
($m_1=\pm m_1'$ and $m_2=\pm m_2'$), or, $c$ is rational and equal to
$c=-(m_1\pm m_1')/(m_2\pm m_2')$, one of the sine terms in the equation
of motion becomes equal to zero (resonance). However, the resonances
do not create any secular term in the {\it solutions} of the equations
since the associated terms simply disappear from the r.h.s of Eqs.(\ref{eqmo}).
Thus, the equations in the next order give $dx_{n+1}/dt$ and $dy_{n+1}/dt$ as
sums of sine terms, and it follows that, if $x^{(n)}$ and $y^{(n)}$ are
sums of cosine terms, $x_{n+1}$ and $y_{n+1}$ are also sums of cosine terms.
Thus, no secular terms appear in the solutions $x^{(n)}$, $y^{(n)}$, $\forall
n=1,\ldots\infty$ since $x^{(1)},y^{(1)}$ are sums of cosine terms.

    \begin{figure}\label{h3rcorb}
    \centering
    \includegraphics[width=\textwidth]{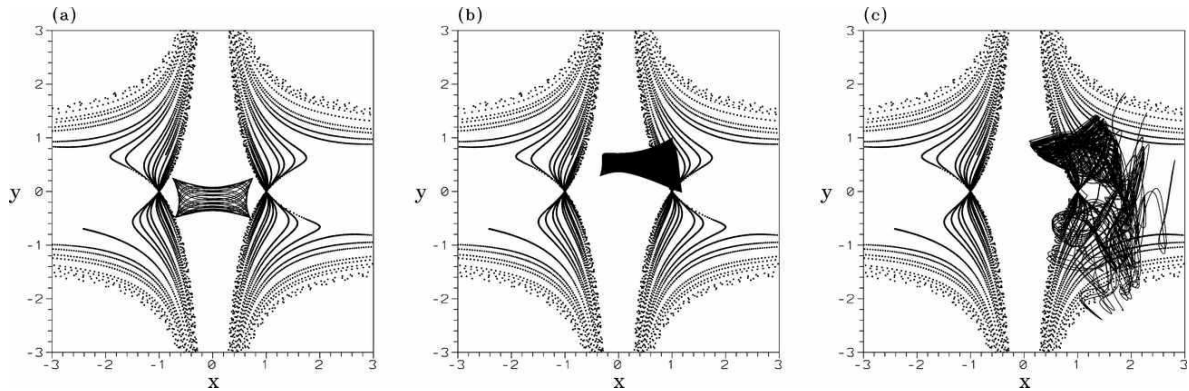}
       \caption{
Three orbits in the equations of motion (\ref{eqmo}) for $a=b=1$,
$c=\sqrt{2}/2$ and $0\leq t\leq 1000$. The initial conditions are
(a) $x(0)=0.75$, $y(0)=0.25$ (regular, not overlapping with the domain of
nodal lines),
(b) $x(0)=y(0)=1$ (regular, partly overlapping with the domain of nodal
lines), and (c) $x(0)=y(0)=1.4$ (chaotic).
}
    \end{figure}

The main characteristics of the regular orbits in the central region can
be understood in terms of the above equations. In particular:

a) The regular orbits are quasi-periodic, i.e., they are given as double
Fourier series with two fundamental frequencies $\nu_1$, $\nu_2$, which
have constant values $\nu_1=1$, $\nu_2=c$. This fact is important because
it implies that there is {\it no} dependence of the frequencies of the
orbits on the amplitudes of the oscillations. This is different from what
is usually encountered in the case of classical nonlinear dynamical systems,
in which the frequencies depend, in general, on the amplitudes. On the
contrary, the quantum mechanical orbits in this system are completely
degenerate with respect to their fundamental frequencies.
    \begin{figure}\label{ha02}
    \centering
    \includegraphics[width=\textwidth]{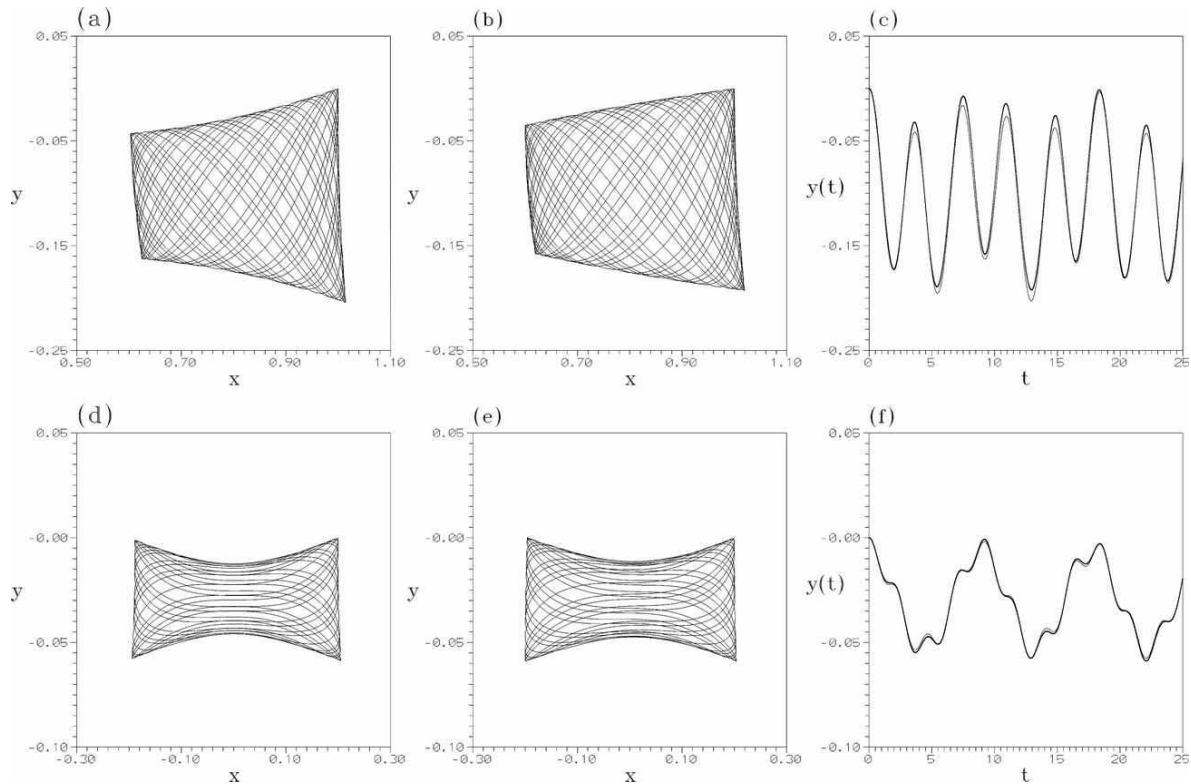}
       \caption{
(a) An orbit with $a=b=0.2$, $c=\sqrt{2}/2$ and initial conditions $x(0)=1$, $y(0)=0$.
(b) The theoretical approximation to the same orbit up to terms of second degree
    in $a$ and $b$, given by Eqs.(\ref{h2solxy0}) and (\ref{h2solyy0}).
(c) The time evolution of $y(t)$ for the same orbit. The numerical and theoretical
    curves almost coincide.
(d) Same as (a) but with initial conditions $x(0)=y(0)=0$.
(e) Same as (b) but for the orbit (d).
(f) Same as (c) but for the orbit (d); the secondary local maxima of $y(t)$
    correspond to deflections of the orbit inside its `box' limit.
}
    \end{figure}

    \begin{figure}\label{orbth}
    \centering
    \includegraphics[width=\textwidth]{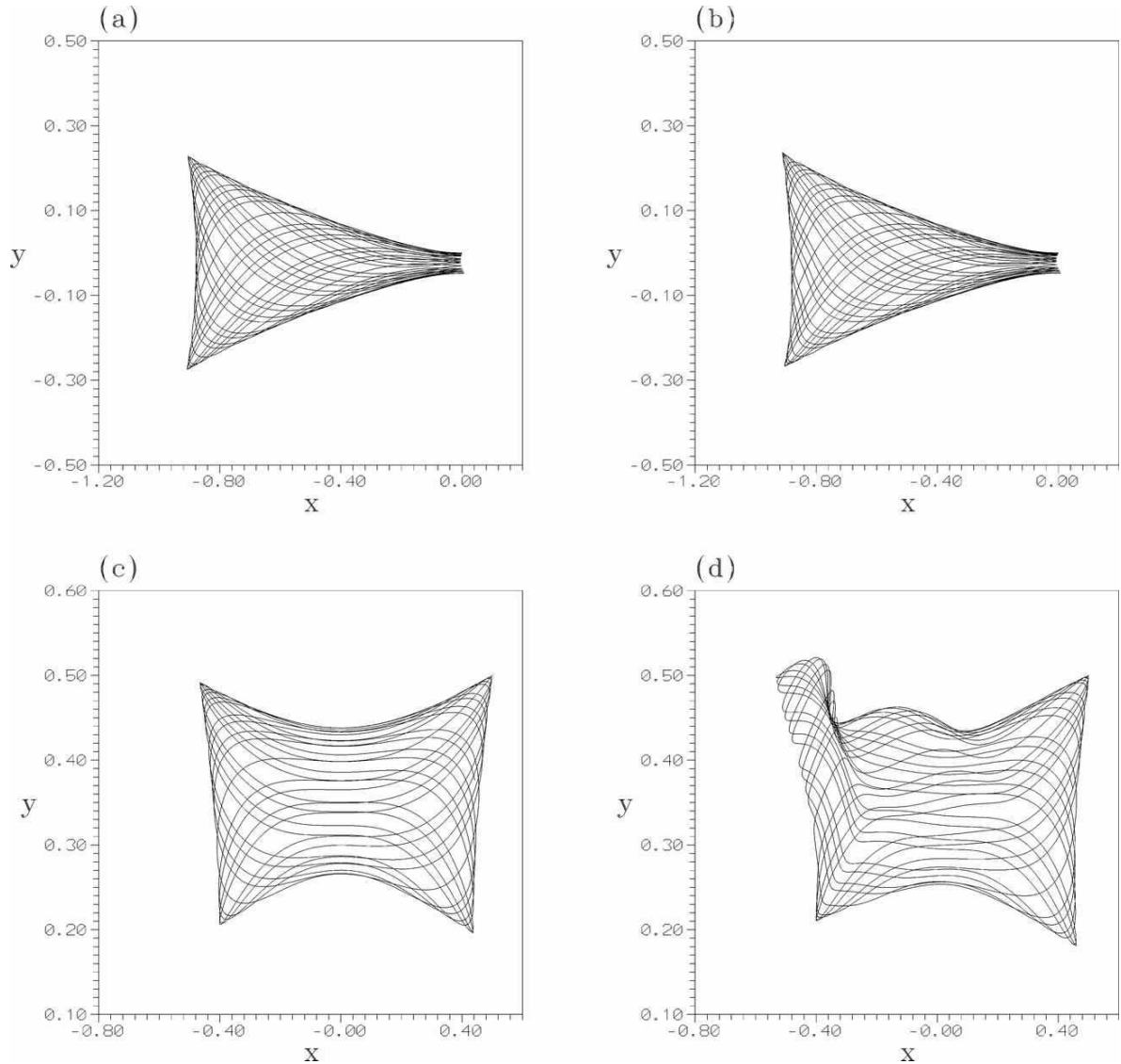}
       \caption{
Numerical versus theoretical orbits for $a=b=0.5$, $c=\sqrt{2}/2$.
(a) Numerical orbit with initial conditions $x(0)=y(0)=0$.
(b) The theoretical approximation of orbit (a) with an expansion up to terms of
    10-th degree in $a$,$b$.
(c) Numerical orbit with initial conditions $x(0)=0.5$, $y(0)=0$.
(d) Same as in (b) but for the orbit (c).
}
    \end{figure}


b) The amplitudes of all the trigonometric terms depend on $a$ and $b$,
but in many terms they depend also on the initial conditions $x_0,y_0$.
The former condition ensures that for $a,b$ sufficiently small a regular
orbit does not overlap with the domain of nodal lines.
This is because the amplitudes of oscillations are $O(a)$ and $O(bc^{1/2})$
in the $x$ and $y$ axes respectively (Eqs.(\ref{h2sol1})), while the minimum
distance of a nodal point from the center is of order $O(|1/a|+|1/b|)$
(section 2). Numerically we find such regular orbits for $a,b$ as high as
$a=b=1$ (Figure 3a). However, the oposite is not true, namely an orbit overlaping
partly with the domain of nodal lines may still be regular (Figure 3b).
In fact, we find numerically that while there is a spatial overlap of the
domains of the orbit and of the nodal lines, the time evolution of both the
orbit and the nodal point is such that their distance is always large (of
order unity). Thus when an orbit enters some region containing nodal lines
the nodal point is far from this region. However, we also find numerically
that if an orbit has significant overlap with the domain of nodal lines,
the orbit is, in general, chaotic (Figure 3c).

c) In the lowest approximation, when $x_0\neq 0$ the theoretical orbits are
`box orbits' (Eqs.(\ref{h2sol1})) like in the classical case. However, when
higher order terms are taken into account, some box orbits develop deflections
internal to the box, while other box orbits have deflections only at the
boxes' limits, depending on the value of $x_0,y_0$. A simple example is
provided by orbits starting on the axis $y_0=0$. Then, the equations
of the orbits are simplified considerably .
We have, up to second order:
\begin{eqnarray}\label{h2solxy0}
x(t)&=&x_0+a[\cos t-1]-{a^2x_0\over 2}[\cos 2t-1] - \nonumber\\
& &{b^2cx_0\over (1+c)^2}[\cos((1+c)t)-1]+
{b^2cx_0\over 4(1+c)^2}[\cos(2(1+c)t)-1]
\end{eqnarray}
\begin{eqnarray}\label{h2solyy0}
y(t)&=&{bcx_0\over 1+c^2}[\cos((1+{1\over c^2})t)-1] \nonumber\\
&-&{abc^{1/2}\over 1+c}[\cos((1+c)t)-1] +
{ab\over 2c^{1/2}}[\cos(ct)-1] \\
&-&{abc^{1/2}\over 2+c}(x_0^2-{1\over 2})[\cos((2+c)t)-1]~~. \nonumber
\end{eqnarray}
In Eq.(\ref{h2solxy0}) the first order term $a[\cos t-1]$ depends only on $a$,
yielding an oscillation of amplitude $2a$, while in Eq.(\ref{h2solyy0}),
the first order term $bc^{1/2}x_0[\cos((1+c)t)-1]/(1+c)$ depends on both $b$
and $x_0$. For all initial conditions with $|x_0|=O(1)$, this term is dominant
over the second order
terms in Eq.(\ref{h2solyy0}). Thus the orbit is a deformed parallelogram,
i.e., it resembles to classical box orbits (Figure 4a,b,c). On the other hand,
if $|x_0|\rightarrow 0$, this term becomes small, while the second order terms,
not depending on $x_0$, become now important. This means that secondary oscillations
of the function $y(t)$ are developed, that may also yield local minima or maxima
besides the main minimum or maximum defined by the first order Fourier component
of (\ref{h2solyy0}). This causes the orbit to develop deflections in the
$y-$ direction inside the box. A numerical example of this behavior is
demonstrated in Figs. 4d,e,f.

Figure 5 shows some further examples of theoretical orbits calculated by the above
series, via a computer program implementing the recursive algorithm up to the 10th
order. As expected in any kind of perturbative series, as $a,b$ increase one needs
higher order terms to obtain a good approximation of the orbits. In fact, as
already analyzed, the approximation depends also on the values of the initial
conditions $x_0,y_0$ which appear in the amplitudes of the trigonometric terms of
Eqs.(\ref{h2sol1}), (\ref{h2sol2x}) and (\ref{h2sol2y}). Thus, when $x_0=y_0=0$, the
theoretical orbit for $a=b=0.5$ (Fig.5a,b) fits well the numerical orbit, while if
the series are truncated at orders well below 10 the agreement is not so good.
On the other hand for
the same amplitudes $a,b$ the fit is not good when $x_0=y_0=0.5$ (Fig.5c,d).
Now, in any perturbation theory the analytical results are precise
up to values of the small parameters smaller than the values for which regular
orbits are found numerically. Thus, in our case we find that for amplitudes larger
than $a=b=0.75$ the approximation of the numerical orbits by series yields
no longer accurate results, despite the fact that we still find numerically many
regular orbits.

\section{The dynamics close to nodal points and the transition to chaos}

\subsection{Phase space structure close to nodal points and the adiabatic
approximation}

Having demonstrated the existence of regular orbits far from the nodal points,
we now examine the motion in the other limit, i.e., close to a nodal point.
Our main remark in the sequel is that the phenomena relevant to the transition
to chaos are unraveled when one considers the passage of the orbits in the
neighborhood of the nodal point in a {\it moving} frame of reference that is
centered at the nodal point. Introducing $u=x-x_0$,
$v=y-y_0$, the equations of motion in the moving frame read:
\begin{eqnarray}\label{equv1}
{du\over dt} &= &-{bc^{1/2}v\sin(1+c)t\over G}-\dot{x}_0 \nonumber\\
{dv\over dt} &= &{bc^{1/2}u\sin(1+c)t-abc^{1/2}u^2\sin ct\over G}-\dot{y}_0
\end{eqnarray}
where $G=G_2+G_3+G_4$ with
\begin{eqnarray}\label{g234}
G_2 &= &{u^2\over x_0^2}-2bc^{1/2}uv\cos(1+c)t+b^2cx_0^2v^2\nonumber\\
G_3 &= &-{2bc^{1/2}\over x_0}u^2v\cos(1+c)t+2b^2cx_0uv^2\\
G_4 &= &b^2cu^2v^2\nonumber
\end{eqnarray}
and $\dot{x}_0,\dot{y}_0$ are found by differentiating $x_0$, $y_0$ with respect
to time in
(Eq.(\ref{nodal}). We then consider the distance $R=\sqrt{u^2+v^2}\equiv\epsilon$ of
an orbit from the nodal point as a small parameter and derive the main characteristics
of the motion by taking expansions of the equations of motion (\ref{equv1}) in powers
of $\epsilon$, i.e., by considering both $u$ and $v$ as $O(R)\equiv O(\epsilon)$.
In polar coordinates $u=R\cos\phi$, $v=R\sin\phi$, the equations (\ref{equv1})
read:
\begin{eqnarray}\label{eqpol}
{dR\over dt} &= &-{abc^{1/2}R^2\cos^2\phi\sin\phi\sin ct\over G}
-\dot{x}_0\cos\phi
-\dot{y}_0\sin\phi \nonumber\\
{d\phi\over dt} &= &{bc^{1/2}\sin(1+c)t-abc^{1/2}R\cos^3\phi\sin ct\over G}
-{1\over R}\dot{y}_0\cos\phi
+{1\over R}\dot{x}_0\sin\phi
\end{eqnarray}
where $G=g_2R^2+g_3R^3+g_4R^4$ and $g_2,g_3,g_4$ are readily specified from
Eq.(\ref{g234}) (see appendix for explicit formulae).

To the leading order ($1/\epsilon^2$), the second of Eqs.(\ref{eqpol}) yields:
\begin{equation}\label{eqphi}
{d\phi\over dt} = {bc^{1/2}\sin(1+c)t\over
R^2\bigg({\cos^2\phi\over x_0^2}
-2bc^{1/2}\cos\phi\sin\phi\cos(1+c)t+b^2cx_0^2\sin^2\phi\bigg) }+...
\end{equation}
Thus, the angular velocity around the nodal point is large (of order
$O(1/\epsilon^2)$) and the period can be made arbitrarily small by approaching
closer and closer to the nodal point. This fact justifies the use of the
{\it adiabatic approximation} in the study of the motions near the nodal point.
That is, at a given initial time $t_0$ we set $t=t_0+\epsilon^2t'$, with
$\epsilon=R_0\equiv R(t_0)$, and find
\begin{equation}\label{eqphi2}
{d\phi\over dt'} = {bc^{1/2}\sin(1+c)t_0\over
{\cos^2\phi\over x_0^2}
-2bc^{1/2}\cos\phi\sin\phi\cos(1+c)t_0+b^2cx_0^2\sin^2\phi }+O(\epsilon^2)~~.
\end{equation}
Now, the denominator in the r.h.s. of Eq.(\ref{eqphi2}) is equal to the
square of the length of one diagonal of the parallelepiped with sides
$|\cos\phi/x_0|$,$|bc^{1/2}x_0\sin\phi|$ forming an angle $(1+c)t_0$,
thus it is always positive. It follows that $d\phi/dt'$ and $d\phi/dt$
have a unique sign during a whole period of $\phi$, which is the same as
the sign of $\sin((1+c)t_0$. That is, at a given time $t_0$, the angular
motions close to the nodal point are all described in the same sense.

In the same approximation we can now determine the form of the integral curves
of the velocity vector field (\ref{equv1}). Dividing the first with the second
of Eqs.(\ref{eqpol}), and setting a constant $t_0$ in the place of $t$ in the r.h.s.
yields the equation of the integral curves, which is of the form:
\begin{equation}\label{dspir}
{dR\over d\phi}=
{A_2(\phi;t_0,x_0,\dot{x}_0,\dot{y_0})R^2+
A_3(\phi;t_0,x_0,\dot{x}_0,\dot{y_0})R^3+...
\over
B_0(\phi;t_0,x_0,\dot{x}_0,\dot{y_0})+
B_1(\phi;t_0,x_0,\dot{x}_0,\dot{y_0})R+...}~~.
\end{equation}
The precise functions $A_i$, $B_i$ are given in the appendix. At this point it
suffices to state that both functions contain only trigonometric terms in $\phi$.
If we expand Eq.(\ref{dspir}) with respect to $R$ we find:
\begin{equation}\label{dspir2}
{dR\over d\phi}=
f_2(\phi;t_0,x_0,\dot{x}_0,\dot{y_0})R^2+
f_3(\phi;t_0,x_0,\dot{x}_0,\dot{y_0})R^3+O(R^4)
\end{equation}
with
$$
f_2={A_2\over B_0},~~ f_3={A_3\over B_0}-{A_2B_1\over B_0^2}~~.
$$
Rescaling the radial distance as $R=\epsilon R'$, with $\epsilon=R_0$,
Eq.(\ref{dspir2}) takes the form:
\begin{equation}\label{dspir3}
{dR'\over d\phi}=\epsilon f_2R'^2
+\epsilon^2f_3R'^3+O(\epsilon^3R'^4)~~.
\end{equation}
This equation satisfies the neccesary conditions for applying the
averaging theorem (e.g. Verhulst 1993). This implies that there is a
near-identity transformation $\bar{R'}=R'+O(\epsilon)$ such that the dynamics
in terms of $\bar{R'}$ is given by:
\begin{equation}\label{dspir3}
{d\bar{R}'\over d\phi}=\epsilon <f_2>\bar{R}'^2
+\epsilon^2<f_3>\bar{R}'^3+O(\epsilon^3\bar{R}'^4)
\end{equation}
where
$$
<f_i>\equiv{1\over 2\pi}\int_0^{2\pi}f_id\phi,~~~i=2,3~~~.
$$
After some algebra (see appendix) we find that $<f_2>=0$ and $<f_3>\neq 0$.
Thus, the equation of the integral curves (back-transformed to non-rescaled
variables) reads finally:
\begin{equation}\label{dspav}
{dR\over d\phi}=<f_3>R^3+...
\end{equation}
where $<f_3>$ depends only on $t_0$ both explicitly and through
$x_0,\dot{x}_0,\dot{y}_0$. The precise form of $<f_3>$, found in the appendix,
reads:
\begin{eqnarray}\label{f3}
<f_3> &= &\Bigg({1+b^2cx_0^4\over 4bc^{1/2}x_0^4\sin(1+c)t_0}\Bigg)\times\nonumber\\
& &\Bigg(
x_0\dot{x}_0
+{\dot{x}_0\dot{y}_0(b^2cx_0^4-1)\over bc^{1/2}\sin(1+c)t_0}
-x_0^2(\dot{x}_0^2-\dot{y}_0^2)\cot(1+c)t_0
\Bigg)~.
\end{eqnarray}
Eq.(\ref{dspav}) admits the solution
\begin{equation}\label{spsol}
R(\phi)={R_0\over \sqrt{1-2R_0^2<f_3>(\phi-\phi_0)}}
\end{equation}
which is a spiral terminating at $R=0$, i.e., at the nodal point, when
$\phi\rightarrow\infty$ (if $<f_3>~<~0$), or $\phi\rightarrow -\infty$ (if
$<f_3>~>~0$). Comparing this to the actual sense of rotation, given by the
sign of $\sin((1+c)t_0$, we can decide whether, as $t$ increases, the orbits
along the spiral recede from or approach to the nodal point. The sense of
rotation changes at times $t_0$ when $\sin(1+c)t_0=0$, i.e., $x_0=0$ and
$y_0=\infty$. On the other hand, the sign of $<f_3>$ changes at times $t_0$
when
$$
x_0\dot{x}_0
+{\dot{x}_0\dot{y}_0(b^2cx_0^4-1)\over bc^{1/2}\sin(1+c)t_0}
-x_0^2(\dot{x}_0^2-\dot{y}_0^2)\cot(1+c)t_0=0~~.
$$
Close to such a time a Hopf bifurcation takes place that is connected to
a change of the character of the nodal point from attractor to repellor,
or vice versa. This phenomenon is analyzed in subsection 4.2 below.

Eq.(\ref{spsol}) is valid only very close to the nodal point. In order to find
the form of the phase flow at larger distances from the nodal point, we look for
stationary points of the flow (\ref{equv1}). The stationary points are given by
non-zero solutions $(u_0,v_0)$ of the system of equations $du/dt=dv/dt=0$. Assuming
$(u_0,v_0)$ small, and keeping terms up to second degree in $u_0$, $v_0$ in
Eqs.(\ref{equv1}), we find the solution:
$$
v_0\simeq \Bigg({\dot{x}_0\over \dot{y}_0}\Bigg)
\Bigg({au_0^2\sin ct_0\over\sin(1+c)t_0}-u_0\Bigg)
$$
which, after replacement in the first of Eqs.(\ref{equv1}), with $du/dt=dv/dt=0$,
yields:
\begin{equation}\label{xpnt}
u_0 \simeq {K(t_0)\over L(t_0,x_0,\dot{x}_0,\dot{y}_0)},~~~
v_0\simeq \Bigg({\dot{x}_0\over \dot{y}_0}\Bigg)
\Bigg({au_0^2\sin ct_0\over\sin(1+c)t_0}-u_0\Bigg)
\end{equation}
where
$$
K=bc^{1/2}\sin(1+c)t_0
$$
and
$$
L=\Bigg(
2\dot{x}_0bc^{1/2}\cos(1+c)t_0+
{\dot{y}_0\over x_0^2}+
{b^2cx_0^2\dot{x}_0^2\over\dot{y}_0}
\Bigg)
+abc^{1/2}\sin ct_0~~.
$$
The stationary point $(u_0,v_0)$ is a saddle (hereafter called `X-point'), with one
positive and one negative real eigenvalues. The reality of eigenvalues follows
immediately by noticing that the variational matrix of (\ref{equv1}) is symmetric
by virtue of the fact that the equations of motion are given by the grad
$\nabla_{u,v}S'$ with $S'=S-\dot{x}_0 u - \dot{y}_0 v$, with $S(u,v,t)$ equal to the
phase of the wavefunction $\psi=Re^{iS}$. Thus, the off-diagonal elements of the
variational matrix are equal, namely:
$$
{\partial\over\partial v}{du\over dt}={\partial^2S'\over\partial v\partial u}=
{\partial^2S'\over\partial u\partial v}={\partial\over\partial u}{dv\over dt}
$$
i.e., the variational matrix is symmetric and its eigenvalues are real.
Furthermore, setting
$a_{11}=\partial(du/dt)/\partial u$,
$a_{12}=\partial(du/dt)/\partial v$,
$a_{21}=\partial(dv/dt)/\partial u$,
$a_{22}=\partial(dv/dt)/\partial v$, the characteristic equation is given by
$$
\lambda^2-(a_{11}+a_{22})\lambda+(a_{11}a_{22}-a_{21}a_{12})=0~~.
$$
To the lowest approximation we find:
$$
a_{11}={bc^{1/2}v\sin(1+c)t_0\over G^2}{\partial G\over\partial u}+...,~~
a_{22}=-{bc^{1/2}u\sin(1+c)t_0\over G^2}{\partial G\over\partial v}+...,~~
$$
$$
a_{12}=a_{21}={1\over 2}
{bc^{1/2}v\sin(1+c)t_0\over G^2}
\Bigg(v{\partial G\over\partial v}-u{\partial G\over\partial u}\Bigg)+...
$$
Thus the roots of the characteristic equation satisfy
\begin{equation}\label{l1l2}
\lambda_1\lambda_2 = -{b^2c\sin^2(1+c)t_0\over 4G^4}\Bigg(
v{\partial G\over\partial v}+u{\partial G\over\partial u}\Bigg)^2+...
\end{equation}
and if we replace $u,v$ in (\ref{l1l2}) by the root $u_0,v_0$, with
$u_0,v_0$ sufficiently small, the product of the eigenvalues is negative.
This means that one eigenvalue is positive and the other negative (except
for degenerate cases in which one eigenvalue is zero).

A further conclusion stems by noticing that, if the X-point has a distance
$d_0=\sqrt{u_0^2+v_0^2}$ from the nodal point, implying that both $u_0$ and
$v_0$ are of order $O(d_0)$, then all the entries $a_{ij}$ of the variational
matrix are of order $O(1/d_0^2)$. It follows that both eigenvalues satisfy:
\begin{equation}\label{lijor}
|\lambda_i|=O({1\over d_0^2}),~~~ i=1,2~~.
\end{equation}
This conclusion is important because it implies that while, as we will see in
the next subsection, chaos is introduced mainly at the approch of the orbits
near an X-point, the contribution of the latter to the positive value of the
Lyapunov characteristic exponent of an orbit is determined by the measure
of the X-point's positive eigenvalue $\lambda$, which, on its turn, is large
when the X-point is close to the nodal point, i.e., when $d_0$ in
Eq.(\ref{lijor}) is small.  This means that the nodal points influence chaos
rather indirectly, that is, the chaotic behavior is actually due to
the X-points, but the effectiveness of the latter depend on their closeness
to the nodal points. Notice that the X-point can approach arbitrarily close
to the nodal point, since the two points collide whenever $K=0$, i.e.,
$\sin(1+c)t_0=0$. This happens whenever the nodal point reaches infinity from
either side of the y-axis. In general, the distance $d_0$ is small when $|y_0|$
is large.

    \begin{figure}\label{nodspir}
    \centering
    \includegraphics[width=10cm]{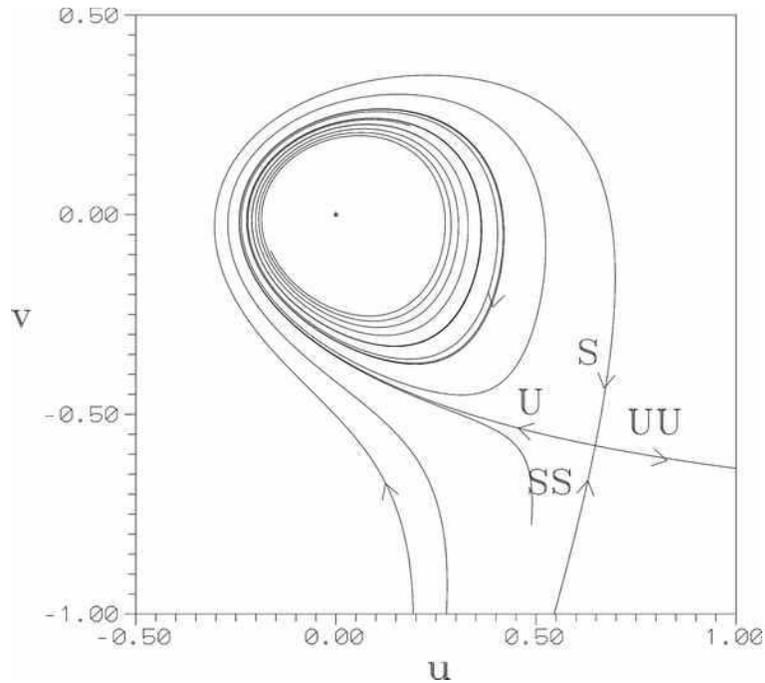}
       \caption{
The nodal point - X point complex on the $(u,v)$ plane, in the adiabatic
approximation when $t_0=10$ and $a=b=1$, $c=\sqrt{2}/2$.
}
    \end{figure}

Figure 6 shows how do the spirals emanating from the nodal point connect to
the invariant manifolds emanating from the X-point in the adiabatic approximation.
This figure is a numerical calculation of all the integral curves emanating
from the X-point, when $a=b=1$, $c=\sqrt{2}/2$ and $t_0=10$.
The X-point is found numerically up to twelve digits by looking
for roots of Eqs.(\ref{equv1}) close to the nodal point by a Newton-Raphson
routine. The numerical solution $(u_0,v_0)$ is then inserted in the expressions
for the matrix elements $a_{ij},i=1,2, j=1,2$, yielding the eigenvalues and
eigenvectors of the variational matrix at $(u_0,v_0)$. Then, we give initial
conditions for $u,v$ on both semi-lines (with respect to $(u_0,v_0)$) determined
by the directions of the two eigenvectors, at a distance $10^{-4}$ from the
X-point. Finally, we integrate numerically the differential equation:
\begin{equation}\label{dudv}
{du\over dv}=f_{uv}(u,v,t_0)
\end{equation}
found by dividing the two equations of (\ref{equv1}), for each of the four
different above sets of initial conditions. This yields numerically the
two branches of the stable and unstable manifolds of the X-point. Clearly,
since all the spirals terminating at the nodal point are described in the
same sense, only one of these four branches can be connected to a spiral
terminating at the nodal point. This branch can always be identified by
comparing the senses of description of the manifolds and of the spiral.
In particular, one of the asymptotic spirals of the nodal point is joined
to one branch of the unstable manifold emanating from the X-point, if the
nodal point is an attractor, or to the stable manifold, if the nodal point
is a repellor. The set of all the integral curves of the flow in the
neighborhood of the nodal point and X-point is hereafter called the
`nodal point - X-point complex'.

Figure 7 shows a comparison of the spatial distribution of the nodal points
(Fig.7a) and of the respective X-points (Fig.7b), on the plane $(x,y)$ when
$a=b=1$, $c=\sqrt{2}/2$ and $t_0$ is in the interval $0\leq t_0\leq 1000$.
The nodal lines and the X-point lines form similar patterns. In particular,
similarly to the nodal points (section 2), the X-points avoid a central
region of the plane $(x,y)$ in which the majority of orbits turn to be
regular (subsection 4.3). Fig.7c shows the modulus of the positive eigenvalue
of the X-point as a function of the distance $d_0$ of the X-point from the
nodal point. We find numerically a power-law scaling $\lambda\propto 1/d^p$
with $p\simeq 1.5$, i.e., less steep than the theoretical scaling given by
Eq.(\ref{lijor}).
From this figure, as well as from Figure 6, in which the distance of the
X-point from the nodal point is about $d_0=0.9$, we deduce that the
results obtained by the previous perturbative analysis are essentially valid
not only at very small distances from the nodal point but also at relatively
large distances (of order $10^{-1}$). At any rate, we always find
that the stationary points of the vector field (\ref{equv1}), as determined
numerically, induce a similar phase-space structure as in Figure 6, i.e., this
structure is general in the model considered.
    \begin{figure}\label{hsq2xpt}
    \centering
    \includegraphics[width=\textwidth]{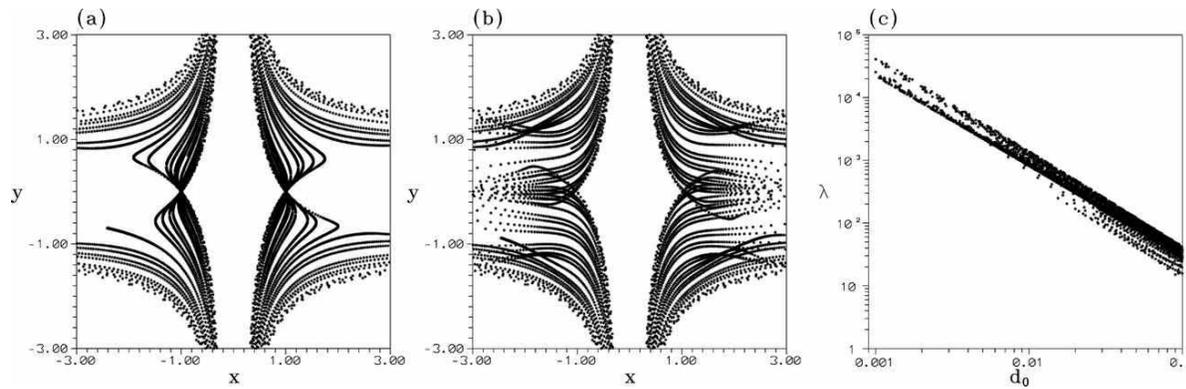}
       \caption{
(a) Nodal lines, (b) X-point lines, and (c) the positive eigenvalue of the X-point
versus the distance $d_0$ of the X-point from the nodal point, when $a=b=1$,
$c=\sqrt{2}/2$.
}
    \end{figure}


\subsection{The exponential sensitivity of the orbits}

In order to understand how does the approach of an orbit to the nodal point -
X-point complex introduce exponential sensitivity of the orbits to the initial
conditions, we consider in detail the successive encounters of two nearby orbits,
with initial separation $10^{-4}$, with this complex, which take place at snapshots
at which the orbits pass close to the complex. To this end, we consider the orbit of
Figure 8a (initial conditions
$x_1(0)=y_1(0)=-1.1$ and $a=b=1$, $c=\sqrt{2}/2$) which has a number of consecutive
encounters with the nodal point - X point complex. This orbit is chaotic, as seen
from the calculation of the `finite time Lyapunov characteristic number'
\begin{equation}\label{chi}
\chi(t)={1\over t}\ln{|\xi(t)|\over|\xi(0)|}
\end{equation}
where $\xi(t)\equiv (dx(t),dy(t))$ is the deviation vector associated with the
orbit $(x_1(t),y_1(t))$, which is found by integrating the variational equations
of motion together with the orginal equations of motion. The limit
$\lim_{t\rightarrow\infty}\chi(t)$ yields the usual Lyapunov characteristic number.
Numerically we find (Fig.8b) that this limit is close to $LCN\simeq 3\times
10^{-2}$. We then consider in detail the growth of deviations from this orbit
by calculating also a nearby orbit $(x_2(t),y_2(t))$ with initial conditions
$x_2(0)=x_1(0)+10^{-4}$, $y_2(0)=y_1(0)$.
    \begin{figure}\label{orb1111}
    \centering
    \includegraphics[width=\textwidth]{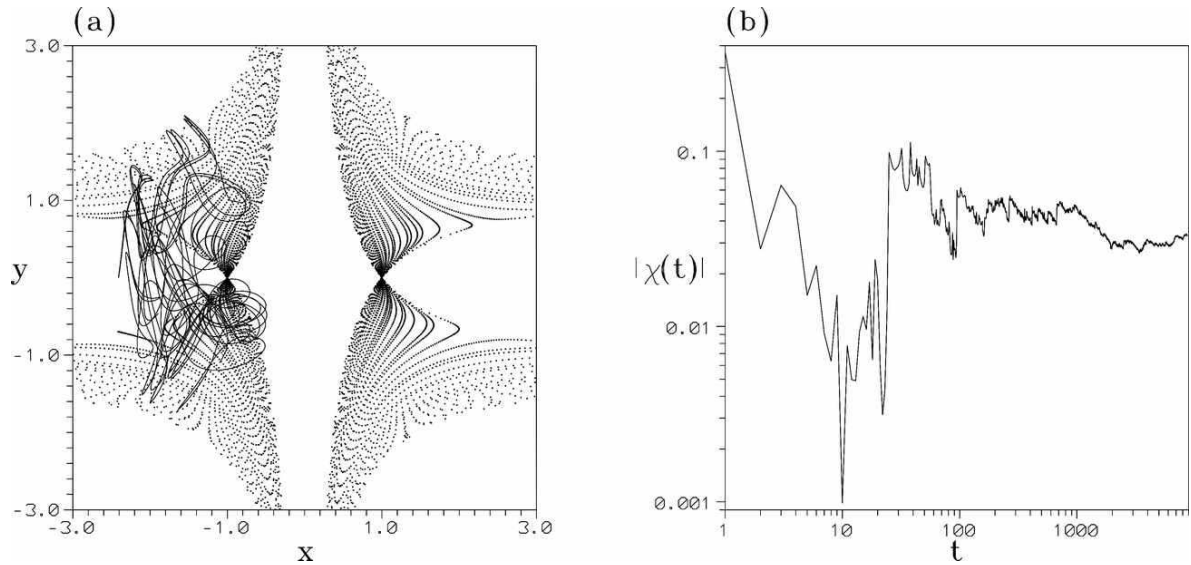}
       \caption{(a)
Chaotic orbit with initial conditions $x(0)=y(0)=-1.1$, and $a=b=1$,
$c=\sqrt{2}/2$. The orbit intersects the nodal lines of the left semiplane
of $(x,y)$. (b) The `finite time' Lyapunov characteristic number $\chi(t)$
for the same orbit. The limiting value is close to $LCN\simeq 0.03$.
}
    \end{figure}


Figure 9a shows the growth in time of the distance
$\Delta S(t)=\bigg((x_1(t)-x_2(t))^2+$ $(y_1(t)-y_2(t))^2\bigg)^{1/2}$
between the two nearby orbits. Clearly, the distance grows in general with
time, but the growth takes place by rather abrupt steps. That is, while the distance
has in general large fluctuations, there are particular times when the distance
$\Delta S$ suddenly grows by jumps of about one order of magnitude (or more).
Thus, the initial distance $\Delta S=10^{-4}$ becomes of order $10^{-3}$ at
a time $t\simeq 25$, then of order $10^{-2}$ at $t\simeq 90$, and finally
of order $10^{-1}$ (reaching even unity) at about $t\simeq 170$. After this
time the distance $\Delta S$ can no longer be considered as small, that is,
the orbits are no longer nearby.

Figure 9b is a close up to the third of the above described jumps, focusing
on the time interval $160\leq t\leq 180$. From this figure it is clear that
there are three encounters of the orbit with the nodal point - X point
complex taking place in the considered time interval. In encounter (I),
the minimum distance of the orbit (1) ($\equiv(x_1(t),y_1(t)$) from the
nodal point is $\epsilon_{min}\simeq 0.4$, and the minimum distance from the
X point is even smaller ($d_{min}\simeq 0.1$). On the contrary, in the
next encounter (II), the minimum distance from the X-point is rather
large ($d_{min}\simeq 0.9$) while the minimum distance from the nodal point
is about the same as in case (I). Finally, in case (III) the minimum
distance from the X-point is small ($d_{min}\simeq 0.2$) while the minimum
distance from the nodal point is now large ($\epsilon_{min}\simeq 1$). Clearly,
the growth of the distance $\Delta S$ mostly takes place during the encounters
(I) and (III) in which the orbits pass closer to the X-point than to the nodal
point. On the other hand, in the case of encounter (II), the growth is smaller
while the orbit approaches closer the nodal point than the X-point.
We conclude that large variations of $\Delta S$ are in general associated with
approaches of the orbits to the X-point rather than to the nodal point.
    \begin{figure}\label{orbdis}
    \centering
    \includegraphics[width=\textwidth]{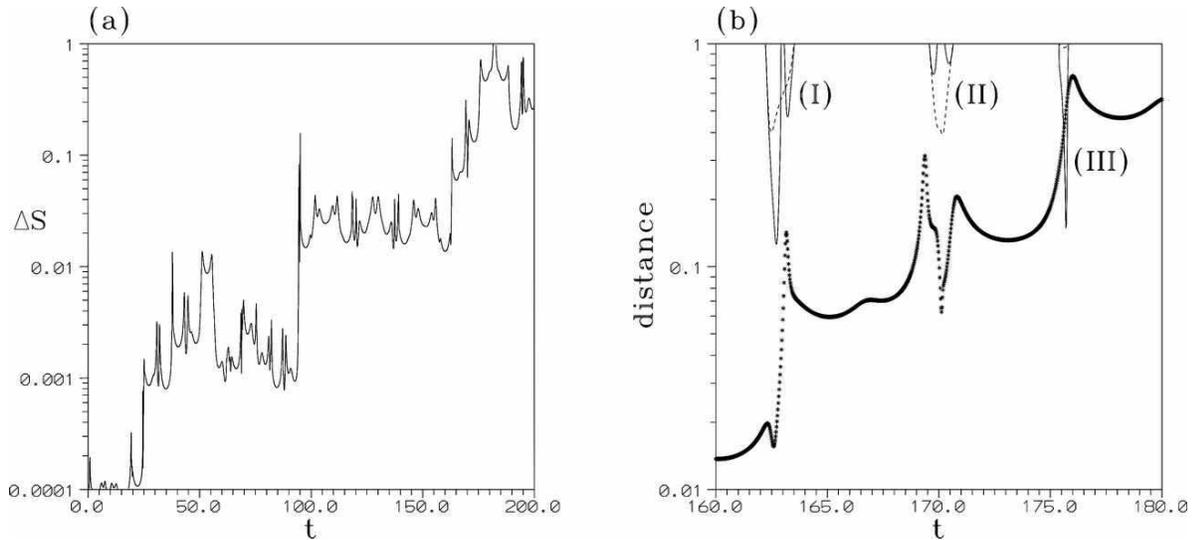}
       \caption{(a) The separation $\Delta S$ of two nearby orbits as a function
of the time $t$ ($a=b=1$, $c=\sqrt{2}/2$ and initial conditions $x_1(0)=-1.1$,
$x_2(0)=x_1(0)+10^{-4}$, $y_1(0)=y_2(0)=-1.1$). (b) A detail of (a) in the time
interval $160\leq t\leq 180$. The line with thick dots gives $\Delta S$.
The solid and dashed lines show the distances
$d(t)$ and $\epsilon(t)$ of the orbit $(x_1(t),y_1(t))$ from the instantaneous
locations of the X-point and of the nodal point respectively, when the latter
are smaller than 1.
}
    \end{figure}

    \begin{figure}\label{deflect}
    \centering
    \includegraphics[width=13.5cm]{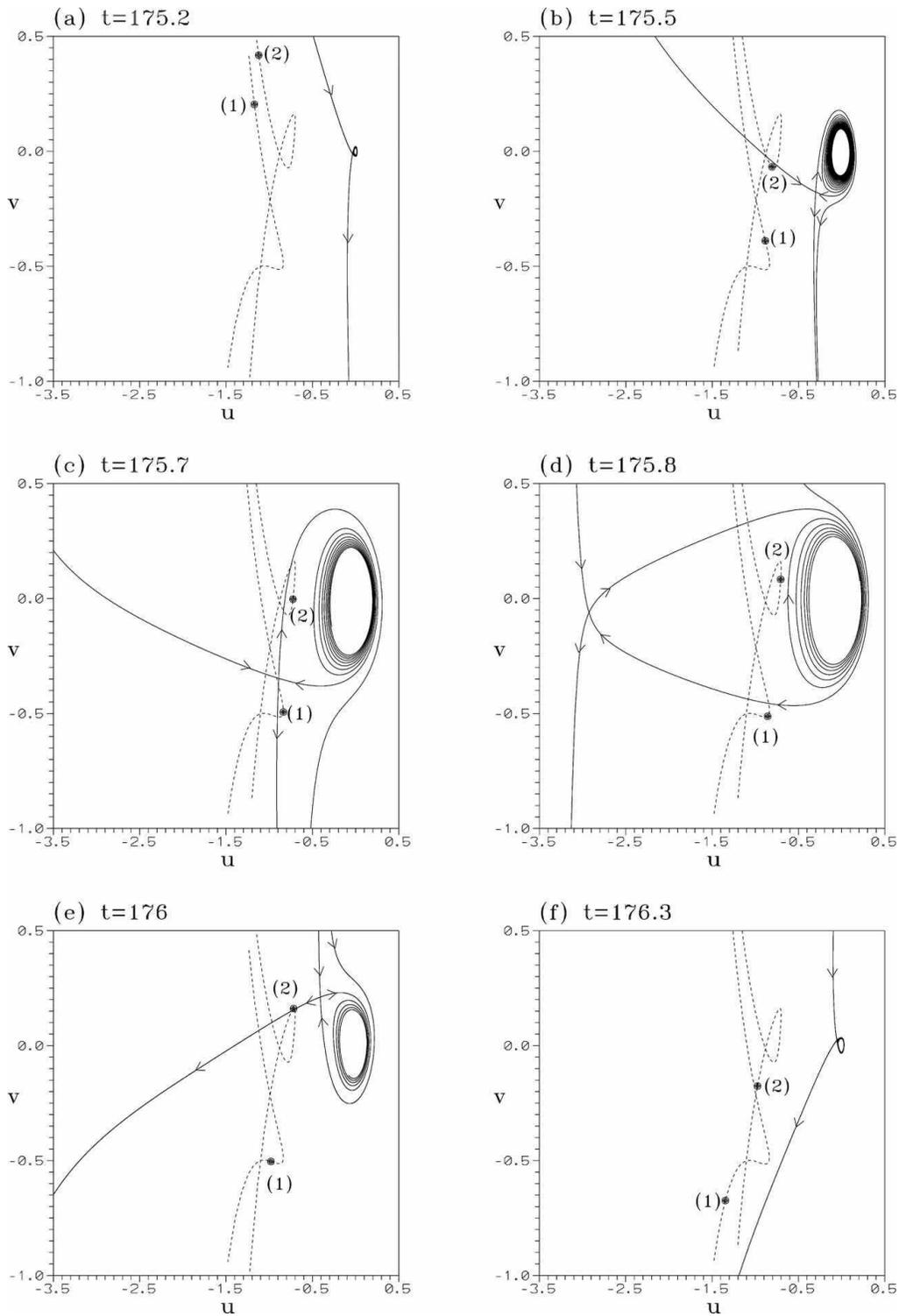}
       \caption{
The encounter event (III) of Figure 9b viewed in detail in the plane
$(u,v)$. The orbits (1)and (2), in the time interval $175\leq t\leq 176.5$,
are shown by dashed lines, while the thick dots indicate the positions of the
orbital points on these lines at the times (a) $t=175.2$, (b) $t=175.5$,
(c) $t=175.7$, (d) $t=175.8$, (e) $t=176$, and (f) $t=176.3$.
The invariant manifolds of the instantaneous X-point - nodal point
complex are plotted for the same times. The main deflection of the
orbits, takes place within the time interval $175.7\leq t\leq 175.8$.
}
    \end{figure}

    \begin{figure}\label{f3t175}
    \centering
    \includegraphics[width=\textwidth]{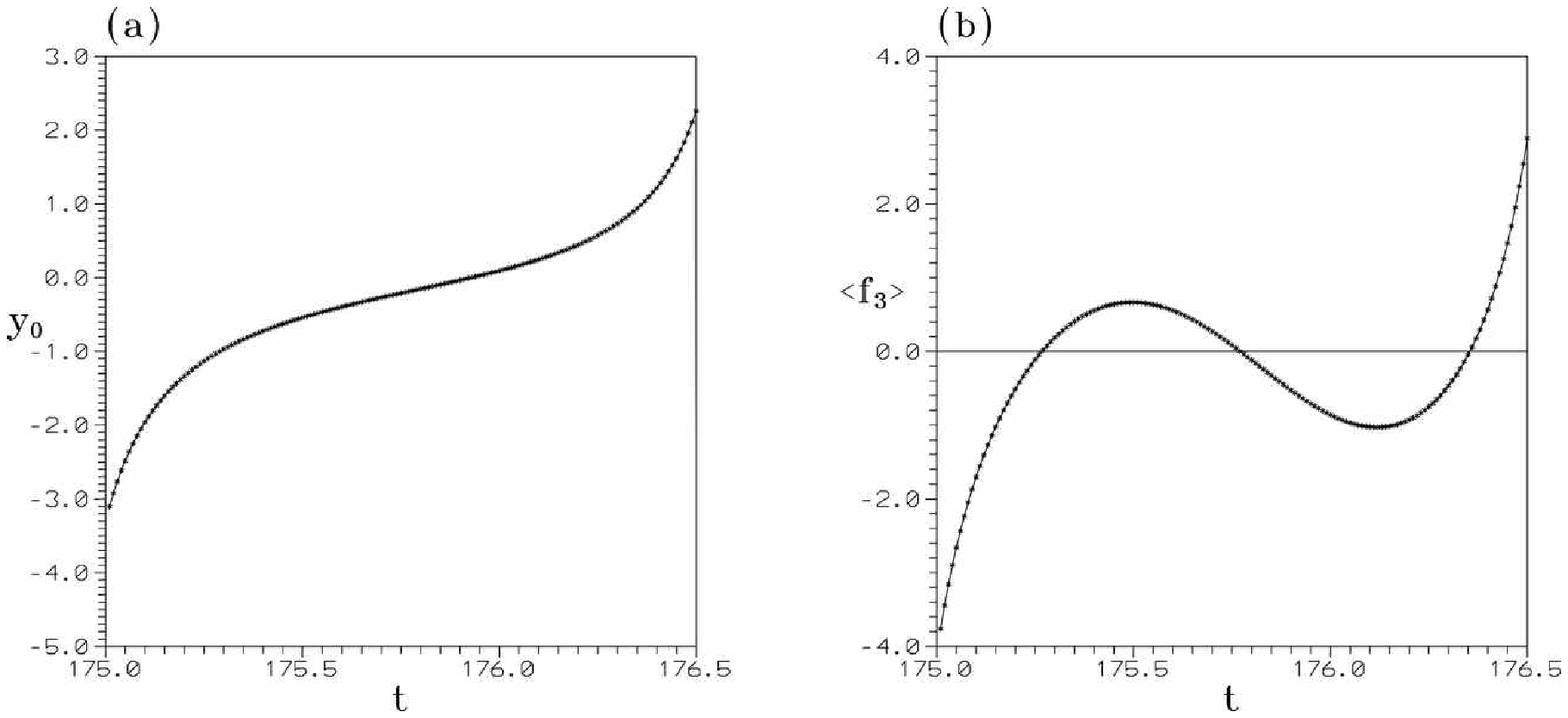}
       \caption{
(a) The nodal point coordinate $y_0(t)$ in the time interval $175\leq t\leq
176.5$. (b) The value of $<f_3>(t_0)$ in the time interval $175\leq t_0\leq 176.5$.
There are three moments in this interval at which $<f_3>=0$. These correspond
to changes in the topological structure of the X-point - nodal point complex.
}
    \end{figure}

Figure 10 shows in detail how does the separation of the orbits take place
during the encounter event (III). The two nearby orbits are shown as dashed
curves, in the time interval $175\leq t\leq 176.5$, in the moving frame of
reference centered at the nodal point. The different frames correspond to
different time snapshots, and the particular positions of the orbital
points $(1)\equiv(u_1,v_1)=(x_1-x_0,y_1-y_0)$, $(2)\equiv(u_2,v_2)=(x_2-x_0,
y_2-y_0)$ at the given snapshot are marked with thick dots. Finally, the
instantaneous stable and unstable asymptotic manifolds emanating from the
X-point (in the adiabatic approximation) are plotted for the time corresponding
to each frame.

We notice that the X-point changes position relative to the nodal point,
which in these frames is always centered at $(u,v)=(0,0)$. In fact, as
already mentioned, the two points collide whenever $\sin(1+c)t=0$ (and then
$y_0=\pm\infty$). There are two consecutive collisions at $t=174.829$ and
$t=176.669$. The time $t=175.2$ (Fig.10a) is close to the first of the above
two collision times and, consequently, the X-point at $t=175.2$ is very close
to the nodal point.  Then, the X-point moves to the left up to about $t=175.8$
(Figs.10b,c,d), and then it returns to the right approaching again the
nodal point (Figs.10e,f). The time $t=176.3$ is close to the second
collision time ($t=176.669$), thus the X-point in Fig.10f comes again
very close to the nodal point.

Now, at the time $t=175.2$ the orbits have a separation of about $\Delta S=0.15$
(Fig.10a). At this time snapshot the orbits move in a nearly parallel way,
and their distances from both the nodal point and X-point are rather large
(of order unity). The orbits move downwards in about the same direction
as indicated by the arrows of the invariant manifolds of the X-point
(in the rest frame $(x,y)$ this means that the nodal point approaches
the orbital points from $y=-\infty$, see also Fig.11a). Furthermore,
as the X-point itself moves from right to left, both orbits approach
to it (Fig.10b, $t=175.5$).

The crucial phenomenon occurs near $t=175.7$ (Fig.10c). Around this time,
the X point crosses a segment joining the orbital points (1) and (2).
The two points have approached the moving X-point at a distance smaller
than $0.2$, but they are on opposite branches of the unstable manifold.
Thus, point (1) moves downwards following one branch of
the unstable manifold of the X point, while point (2) moves upwards
following the other branch of the same manifold. This causes a abrupt
growth of the distance of the two points by a factor $\simeq 3$.

An important change in the topological structure of the invariant
manifolds, that influences the orbits, takes place between $t=175.7$
(Fig.10c) and $t=175.8$ (Fig.10d). Namely, at $t=175.7$ (Fig.10c) the
spiral terminating at the nodal point is connected to one branch of
the stable manifold of the X-point, while at $t=175.8$ (Fig.10d) it
is connected to one branch of the unstable manifold of the X-point.
This transition takes place via a Hopf bifurcation which is described
in detail below. At any rate, at $t=175.8$ (Fig.10d) the X-point has
moved to the left, far from the nodal point, and orbit (1) is close to
the stable manifold of the X-point. Thus, orbit (1) is deflected to
the left, while orbit (2) continues slowly upwards.
    \begin{figure}\label{bifurc}
    \centering
    \includegraphics[width=\textwidth]{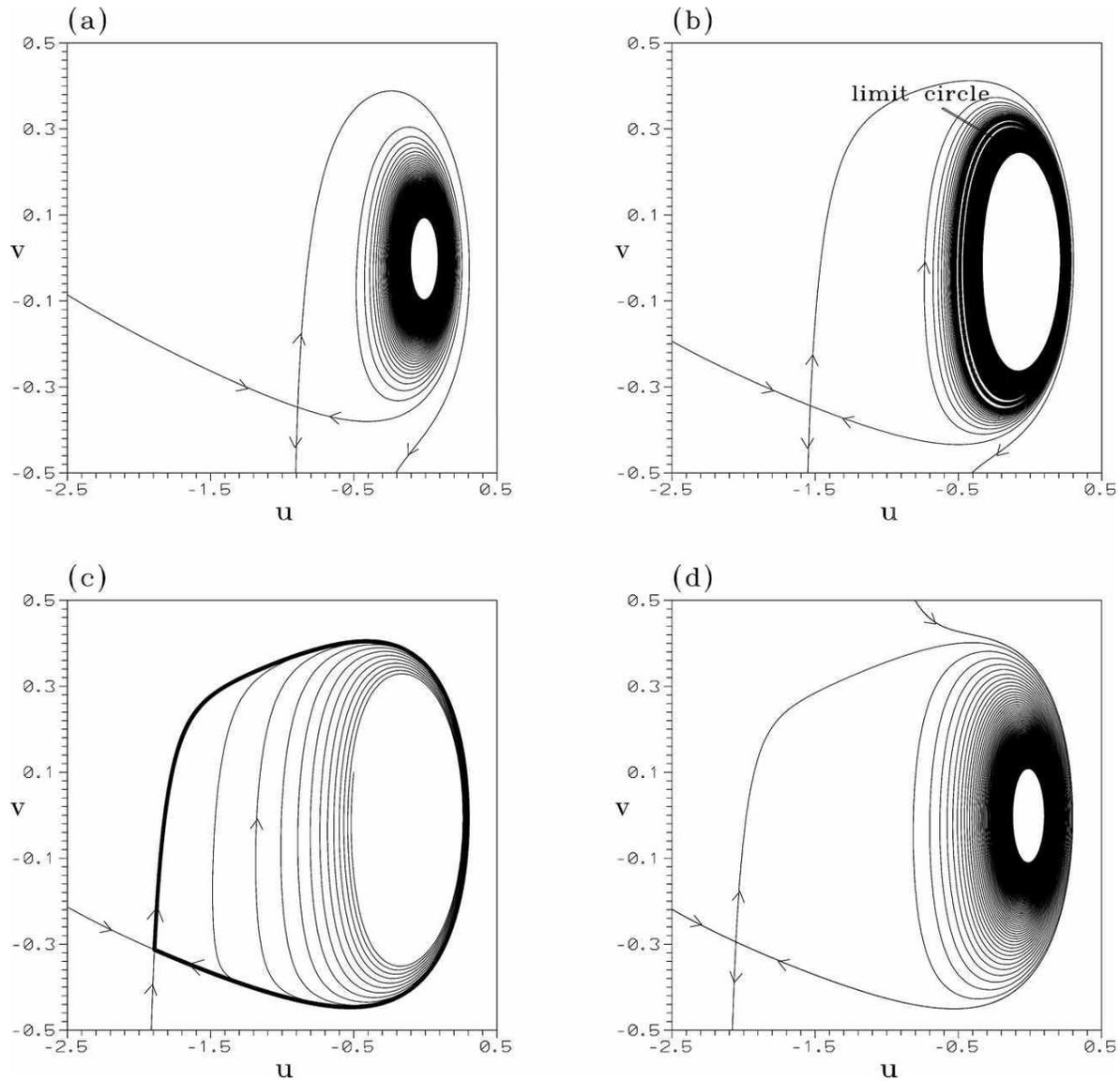}
       \caption{
The instantaneous flow chart of the X-point - nodal point
complex at the times (a) $t_0=176.7$, (b) $t=176.76$, (c) $t=176.7750938$,
(d) $t=175.78$. A Hopf bifurcation taking place near the value $t=176.72$
(at which $<f_3>=0$ in figure 11), leads to the formation of
a limit circle (repellor), shown in (b), which disappears after reaching
the separatrix of (c).
}
    \end{figure}


Finally, a little later ($t=176$, Fig.10e), the X-point returns close to
the nodal point so that point (2) comes very close to the unstable manifold
of the X-point. This causes a deflection of orbit (2) to the left, while
orbit (1), although far from the X-point - nodal point complex,
moves also to the left, following the general direction
of motion indicated by the unstable manifold of the X-point. Finally,
at $t=176.3$ (Fig.10f) both orbital points are far from the X-point
- nodal point complex, but they are relatively close to the unstable
manifold of the X-point in a downwards direction. From there on
both orbits move in a nearly parallel way until the next encounter
event which occurs much later. The overall growth of
the distance of the two orbits by a factor 3 in a time $\Delta t=
176.3-175.2=1.1$ corresponds to an exponential growth rate
$\ln 3/1.1\simeq 1$ in this time interval. This is much larger than
the average exponential growth rate (=LCN$\simeq 0.03$) for the same
orbit within a much longer time interval. This fact justifies the statement
that the growth of $\Delta S$ is by abrupt jumps, which take place during
local (in space and time) encounters with the nodal point - X-point complex.

The topological transition in the phase space structure taking place
between $t=175.7$ (Fig.10c) and $t=175.8$ (Fig.10d) is due to a Hopf
bifurcation taking place between these two times as a result of the
fact that the value of $<f_3>$ (Eq.(\ref{f3})) changes sign (Fig.11b),
while the sign of $d\phi/dt$ remains constant (because there is no
change in the sign of $\sin(1+c)t$, i.e., no transition of $y_0$ to infinity,
in the same interval, Fig.11a). By virtue of Eq.(\ref{spsol}), the change
of the sign of $<f_3>$ at a time between $t_0=175.7$ and $t_0=175.8$ implies
that the nodal point turns from repellor to attractor. The precise time when
this happens depends on higher order terms in the development of the equations
of motion around the nodal point, but it is nevertheless close to the time
when the term depending on $<f_3>$ becomes equal to zero. Numerically,
we find the bifurcation to take place near $t_0=175.75$. Before this
time (e.g. at $t_0=175.7$, Fig.12a) the nodal point is a repellor,
and a spiral emanating from it joins the stable manifold of the X-point.
On the other hand, a little after this time ($t_0=175.76$, Fig.12b),
the nodal point has become an attractor, while the stability character
of the X-point does not change appreciably. Thus, between the nodal
point and X-point there is now a limit circle which acts as a repellor,
i.e., the orbits on both sides of the circle move away from it
on spirals either terminating at the nodal point or moving towards the
X-point. In the latter case a spiral either joins the stable manifold
terminating at the X-point or continues downwards, away from the X-point,
in the channel formed between the two branches of the unstable manifold
of the X-point (Fig.12b). As $t_0$ increases the limit circle moves outwards
approaching the invariant manifolds of the X-point. At a critical time
$t_0=175.7750938$ (Fig.12c) the limit circle coincides with the invariant
manifolds of the X-point. At still larger times ($t=175.78$, Fig.12d),
the limit circle disappears and one branch of the unstable manifold of
the X-point continues now as a spiral terminating at the nodal point.

Figure 13 shows how does the effect of a close encounter of an orbit
with the nodal point - X point complex shows up in the time evolution of the
deviations $\xi(t)$ as given by solving the variational equations of motion
together with the equations of motion for one orbit. The abrupt jumps in the
length of the deviation vector $\mathbf{\xi}=(dx,dy)\equiv(du,dv)$ (Fig.13a)
are associated with passages of the orbit close to the nodal point - X-point
complex. In order to obtain quantitative estimates of the exponential growth
of deviations at successive encounter events, we proceed as follows: The time
span of the total run of an orbit is split in short windows of width $\Delta t=0.1$
(the timestep of the numerical integration is variable and much shorter than this
window, i.e., $dt\leq 10^{-5}$). In each time window, an average `stretching number'
(Voglis and Contopoulos 1994) is calculated according to:
\begin{equation}\label{alfa}
a_i={1\over\Delta t}\ln|{\xi(t_i+\Delta t)\over\xi(t_i)}|
\end{equation}
where $t_i=(i-1)\times\Delta t$ is the initial time of the i-th window,
and $\xi(t)$ is the length
of the deviation vector $\mathbf{\xi}(t)$ at the time $t$. This quantity
characterizes the local growth rate of deviations, while the average value
of all the stretching numbers yields the `finite time Lyapunov characteristic
number'
\begin{equation}\label{chi}
\chi(t)={1\over N\Delta t}\sum_{i=1}^N\ln|{\xi(t_i+\Delta t)\over\xi(t_i)}| =
{1\over t}\ln|{\xi(t)\over\xi(0)}|~~.
\end{equation}
(The limit $\lim_{t\rightarrow\infty}\chi(t)$ defines the Lyapunov characteristic
number of the orbit). In the same time windows we also store the values
of the minimum distance of the orbit to the nodal point $\epsilon_{min}$,
and to the X-point $d_{min}$, as well as the minimum distance between
the X-point and nodal point $d_{0,min}$. These three minimum values are not
occurring at precisely the same times within a given time window, however
the width of the window $\Delta t=0.1$ itself is small enough so that the
occurrences of the minimum values can be considered as nearly simultaneous.

Figure 13b then shows the main result. The distances that an orbit reaches
from either the nodal point or the X-point, for all the 10000 time windows
in a total time span $0\leq t\leq 1000$, are grouped in bins of width $\delta=
2.5\times 10^{-2}$. The abscissa in Fig.13b gives the median value of each bin,
which is the same for either the distance $d$ from the X-point or $\epsilon$
from the nodal point. A weighted average value $\bar{a}$ of the stretching number
is then calculated in each bin, by summing the values of the stretching numbers
which appear during all the passages of the orbit at distances $d\pm\delta/2$ from
the X-point, or $\epsilon\pm\delta/2$ from the nodal point, and dividing by the
total number of values of the stretching number in the sample. Clearly, when
the stretching numbers are grouped with respect to the various distances $d$
reached from the X-point, the average stretching number $\bar{a}(d)$ is positive
for all distances $d\leq 0.25$, while for $d>0.25$, $\bar{a}(d)$ fluctuates between
positive and negative values, showing nevertheless a preference for positive values.
Such a preference reflects the hyperbolic dynamics induced on the orbits by their
approaches to the X-point. That is, the general solution of the variational
equations close to the X-point contains terms growing exponentially and other
terms decaying exponentially. However, the growing terms prevail as the time $t$
increases. Thus, the overall average stretching number $<\bar{a}>$ after many
encounters of an orbit with the X-point turns to be positive.

On the other hand,
when the stretching numbers are grouped with respect to the distance
$\epsilon$ of an orbit from the nodal point, the average stretching number
$\bar{a}(\epsilon)$ (dashed curve in Fig.13b) shows  no clear preference towards
positive or negative values when $\epsilon$ is small ($\epsilon<0.25$),
while a preference towards positive values of $\bar{a}$ appears when $\epsilon>0.25$.
This agrees with Fig.13a, or Fig.9b, which show that the growth of deviations
occurs mainly during encounters (I) and (III), during which the minimum distance
of the orbit from the X-point is smaller than the distance from the nodal point.
    \begin{figure}\label{ast1111}
    \centering
    \includegraphics[width=\textwidth]{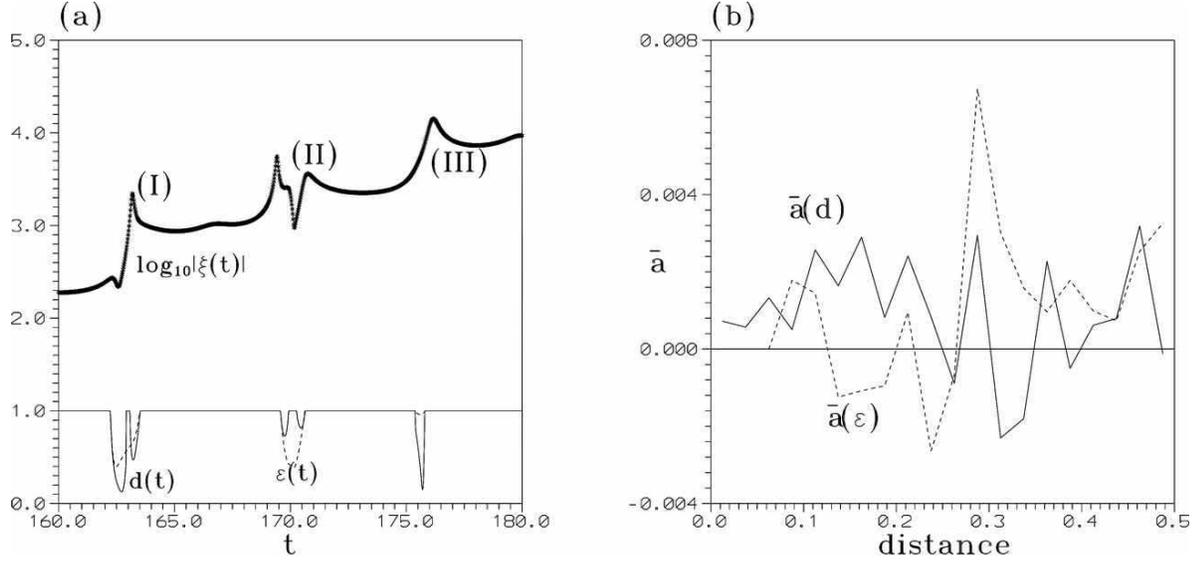}
       \caption{
(a) the growth of the length of the deviation vector $\xi(t)$ for the
same orbit $(x_1(t),y_1(t))$ as in Figure 9b. The lower solid
and dashed lines are the same as in Fig.9b.
(b) The mean stretching number $\bar{a}$ as a function of the distance
from the X-point $\bar{a}(d)$, or from the nodal point $\bar{a}(\epsilon)$,
for the same orbit and time up to $t=1000$.
For each value of $d$ or $\epsilon$, the value of $\bar{a}$ is calculated
by taking the sum of all the stretching numbers (Eq.(\ref{alfa}) with
$\Delta t=0.1$) occuring at passages of the orbit at a distance
$d\pm \delta/2$ from the X-point and $\epsilon\pm\delta/2$ from
the nodal point respectively, with $\delta=2.5\times 10^{-2}$.
}
    \end{figure}


\section{Discussion}
The role of the nodal points of the wavefunction (sometimes called `quantum vortices')
in determining the main features of the `hydrodynamical' probability flow has
been studied in a number of different quantum systems (see Wyatt (2005) and
references therein). Here we refer only to studies related to our own, i.e.,
to the appearance of chaos due to nodal points (quantum vortices).

Berry (2005) studied the flow lines of a general time-independent complex scalar
field $\psi(x,y)$ and found that they typically spiral in or out of a `phase
vortex'. He also found the stationary points of this flow which can be either
elliptic or hyperbolic. Berry explicitly excludes a spiral flow near the
vortices of the Schr\"{o}dinger field because the quantum
mechanical current satisfies $\nabla\cdot\mathbf{j}=0$. This is precisely
what happens in our case if one considers the flow lines in the {\it rest
frame} $(x,y)$ rather than a frame moving together with the nodal points.
In the rest frame, the flow integral curves (found by dividing by parts
Eqs.(\ref{eqmo})) are given by the differential equation:
\begin{equation}\label{velfield}
{dy\over dx} = {bc^{1/2}x\big(ax\sin ct + \sin(1+c)t\big) \over
a\sin t + bc^{1/2}y\sin(1+c)t}~~.
\end{equation}
Keeping the time $t$ frozen in the r.h.s., Eq.(\ref{velfield}) yields the integral
curves
\begin{equation}\label{velcurve}
a\sin t y + bc^{1/2}\sin(1+c)t {y^2\over 2}
- abc^{1/2}\sin ct{x^3\over 3} - bc^{1/2}\sin(1+c)t{x^2\over 2} = C
\end{equation}
The critical points of (\ref{velcurve}) are given by the solutions of
$\nabla C=0$. There are two solutions:
\begin{eqnarray}\label{crit}
(x_1,y_1) &= &(-{\sin(1+c)t\over a\sin ct},-{a\sin t\over bc^{1/2}\sin (1+c)t})
\equiv(x_0,y_0) \nonumber\\
(x_2,y_2) &= &(0,-{a\sin t\over bc^{1/2}\sin (1+c)t})\equiv(0,y_0)~~.
\end{eqnarray}
The first critical point $(x_1,y_1)$ coincides with the nodal point.
The eigenvalues of the Hessian matrix of $C$ at $(x_0,y_0)$ are given by
$\lambda_{1,2} = \pm ibc^{1/2}\sin(1+c)t$, thus they are imaginary at any time
$t$, implying that the integral curves of the velocity field in the neighborhood
of the nodal point are approximately ellipses centered at $(x_0,y_0)$ at any time
$t$ except $t_c=2k\pi/(1+c)$ or $t_c=(2k+1)\pi/(1+c)$ with $k$ integer. The
difference with respect to the approximation of Eq.(\ref{spsol}) is that
in the moving frame of reference the instantaneous flow lines form spirals
if $<f_3>\neq 0$, i.e., if $\dot{x}_0\neq 0$ or $\dot{y}_0\neq 0$,
that is the spirals appear (in the moving frame) only because the velocity of
the nodal point is non-zero. Furthermore, in our analysis, the eigenvalues of the
X-point in the nodal point - X-point complex scale as an inverse power of the
distance of the X-point from the
nodal point. On the contrary, the eigenvalues of the second critical point
$(x_2,y_2)$ of Eqs.(\ref{crit}) are given by $\lambda_{1,2} = \pm bc^{1/2}\sin(1+c)t$, i.e. they are bounded by quantities
of order $a$ or $b$. Thus, despite the fact that $(x_2,y_2)$ represents also a saddle
point in the rest frame of motion in the adiabatic approximation, its influence to the
dynamics is not so important when $a$ and $b$ are of order unity. In fact, this
point is always attached to the y-axis, so that it can only influence the deviation
vectors at times when the orbits come close to this axis. This is at variance with
the numerical results showing that positive stretching numbers are introduced when
the orbits approach the nodal point - X-point complex at arbitrary locations
within the plane $(x,y)$.

Wisniacki and Pujals (2005) studied the Bohmian orbits in an example
similar to ours, namely the superposition of three stationary states of
the double harmonic oscilator, when, however, the latter is isotropic
($c=1$) and the ratios of the probability amplitudes of the states $|\Psi_{10}>$
and $|\Psi_{11}>$ with that of $|\Psi_{00}>$ are complex. In the case $c=1$ there
is only one fundamental
frequency of the time-dependent trigonometric terms of the equations
of motions. This allows one to obtain stroboscopic plots of the orbits,
i.e., Poincar\'{e} surfaces of section. Wisniacki et al. pointed out
that it is the {\it motion} of the nodal point which generates chaos.
However, their mechanism of introduction of chaos is different from ours.
Namely, in the case of Wisniacki et al. the motion of the nodal point
generates a saddle point on the surface of section, but the surface of
section is area preserving, and the transverse intersections between
the stable and unstable invariant manifolds of the saddle point generate
homoclinic chaos. The most important difference is that their mechanism
can be applied only in resonant cases while our mechanism applies to
general non-resonant cases.

A similar example was studied by Makowski et al. (2000). Both results correspond
to the case of a complex ratio $a/b$ and $c=1$. On the other hand, if $c$ is rational
but $a/b$ is real, all the orbits are periodic and neutrally stable, thus there is no
chaos at all (some examples of seemingly chaotic orbits in a similar model given by
Konkel and Makowski (1998), for a rational value of $c$, are just due to numerical
errors caused by the stiffness of the equations of motion close to the X-point.
In order to obtain the correct orbits, which are periodic and not chaotic, we had
to use a program in Mathematica with an accuracy of 50 digits!).

Falsaperla and Fonte (2003) studied the orbits near nodal lines in a
3D model. In that case the orbits describe helical motions around the
nodal line (called by these authors `spirals') while the projections
of the motion on a $z=$constant surface are ellipses. These authors
point out that the nodal lines ``regularize the motion'' in their
neighborhood, and only orbits not following a definite nodal line
are ``intermittent chaotic''. They furthermore find too that the
period of rotation along the helix is of order $O(\epsilon^2)$, where
$\epsilon$ in this case is the distance from the nodal line. While a
detailed comparison of 2D and 3D models is necessary, our analysis
above shows essentially why the growth of the deviation vectors
is intermittent, i.e., it takes place by abrupt steps whenever an orbit
approaches the nodal point - X-point complex. Furthermore, we also explain
why the appearance of chaos is {\it not} strictly correlated with very
close approaches to the nodal point (in such cases the orbits simply
spiral around the nodal point), but it occurs mainly when the orbit
approaches closely the X-point associated with the nodal point. This
happens even if the latter has a distance from the nodal point which is
much larger than the distance of the orbit from the X-point.

Wu and Sprung (1999) gave plots (their Fig.4) of the probability flow, which
resemble our Figure 7. However, this resemblence is only due to a `phase
mixing' phenomenon taking place in the rest frame near the nodal point.
Namely, because the rotation frequency depends on the distance from the
nodal point as $1/\epsilon^2$, a fluid element of some thickness approaching
the nodal point forms a number of windings around the nodal point due to
the differential rotation of the orbits included in its area. These
windings give the impression of forming a spiral pattern, which is,
however, only apparent, namely the windings are limited by an inner
circle (or ellipse) due to orbits of the fluid element closest to the
nodal point (this is different from our limit circle of Fig.12). On the other
hand, in earlier works the same authors found spiral motions around quantum
vortices in the {\it rest frame} of motion, when the quantum system is subject
to a vector potential in addition to a scalar potential (Wu and Sprung 1994).
The plots shown in that work are similar to ours, although the similarity in
the topological structure of the phase space is due to the presence of an extra
term in the equations of motion due to the vector potential, rather than to the
motion of the nodal point.

Finally, Frisk (1997) and Wisniacki et al. (2006) noticed that the transition
to chaos is enhanced when there are many co-existing vortices influencing
the orbits. In that case it is necessary to consider the connections of
the invariant manifolds emanating from the stationary points of one
vortex with those of other vortices and see how this can increase chaos
and transport phenomena in configuration space. This problem is
proposed for future study.

\section{Conclusions}
In summary, the main conclusions of our study are the following:

1) In a simple quantum system consisting of the superposition of three eigenstates
of the double harmonic oscillator potential, in which there is one nodal point
of the wavefunction travelling in the plane $(x,y)$, we proved the existence of
domains of this plane which are free of nodal points. Bohmian orbits in these
domains, as well as orbits slightly overlapping with the nodal lines, are regular.

2) In the central domain devoid of nodal lines the equations of motion admit
expansions in powers of
the probability amplitudes of the eigenstates, yielding the trajectories as
double Fourier series in the fundamental frequencies of the system. These
series represent theoretical orbits which are, by definition, regular.
We show the agreement of the theoretical and numerical orbits in this domain
for sufficiently small values of the probability amplitudes and sufficiently
high order of the expansions, and use this to explain the morphological
characteristics of the regular orbits.

3) Close to the nodal points, we use different expansions, in powers of the
distance from the nodal point, in order to unravel the dynamics. The angular
frequency of motion in a $\epsilon-$ neighborhood of the nodal point is of
order $O(1/\epsilon^2)$. This justifies the use of the adiabatic approximation.
In the fixed frame $(x,y)$ the flow lines close to the nodal point are ellipses.
However, in a {\it moving} frame attached to the nodal point, the flow lines
are spirals terminating at the nodal point. The temporary sense of description
of the spirals by the orbits is unique, i.e., at a given time the nodal point
is either an attractor or a repellor. Furthermore, at a finite distance from the
nodal point there is a saddle stationary point of the flow with one asymptotic
manifold joining the spiral and the other three extending to infinity.

4) The eigenvalues of the X-point scale as $1/d_0^p$ where $d_0$ is the distance
of the X-point from the nodal point and $p\simeq 2$. Furthermore, the
distance $d_0$ can be arbitrarily small, i.e., there are collisions of the
X- and nodal points. We show that the orbits approaching close to the
X-point exhibit exponential growth of their deviations, i.e., they are
chaotic. In all numerical examples we find that chaos is associated with
the approach of the orbits to the X-point, which, however, is only
guaranteed when we have a moving nodal point. Furthermore the chaotic
influence of the X-point on the orbits is strongest when the X-point is
closest to the nodal point. Thus, the nodal point indirectly influences the
transition of Bohm's trajectories from order to chaos via the above mechanism. \\
\\

\clearpage
\section{Appendix: Calculation of $<f_3>$}

The precise form of Eq.(\ref{dspir}) reads:
\begin{equation}\label{dspirpre}
{dR\over d\phi}=\frac
{-AR^2-\dot{x}_0G\cos\phi-\dot{y}_0G\sin\phi}
{B-CR-\dot{y}_0{G\over R}\cos\phi+\dot{x}_0{G\over R}\sin\phi}
\end{equation}
where
$$
A=abc^{1/2}\cos^2\phi\sin\phi\sin ct,~~B=bc^{1/2}\sin(1+c)t,~~
C=abc^{1/2}\cos^3\phi\sin ct
$$
and $G=g_2R^2+g_3R^3+g_4R^4$ with
$$
g_2={\cos^2\phi\over x_0^2}-2bc^{1/2}\sin\phi\cos\phi\cos(1+c)t+b^2cx_0^2\sin^2\phi
$$
$$
g_3=-{2bc^{1/2}\over x_0}\cos(1+c)t\cos^2\phi\sin\phi+2b^2cx_0\cos\phi\sin^2\phi
$$
$$
g_4=b^2c\cos^2\phi\sin^2\phi~~.
$$
Expanding Eq.(\ref{dspirpre}) in ascending powers of $R$ we find:
\begin{eqnarray}\label{dspirser}
{dR\over d\phi} &=
&\Bigg(-AR^2
-\dot{x}_0g_2\cos\phi R^2-\dot{x}_0g_3\cos\phi R^3
-\dot{y}_0g_2\sin\phi R^2-\dot{y}_0g_3\sin\phi R^3\Bigg) \nonumber\\
&\times &{1\over B}\Bigg(1+{c\over B}R+\dot{y}_0{g_2\over B}\cos\phi R
-\dot{x}_0{g_2\over B}\sin\phi R\Bigg)+ \ldots
\end{eqnarray}
Averaging Eq.(\ref{dspirser}) we find:
\begin{equation}
{d\bar{R}\over d\phi}={1\over 2\pi}\int_{0}^{2\pi}{dR\over d\phi}d\phi=<f_2>R^2+
<f_3>R^3+...
\end{equation}
where
$$
f_2={1\over B}(-A-\dot{x}_0g_2\cos\phi-\dot{y}_0g_2\sin\phi)
$$
$$
f_3={1\over B}(-A-\dot{x}_0g_2\cos\phi-\dot{y}_0g_2\sin\phi)
\times
({c\over B}+\dot{y}_0{g_2\over B}\cos\phi
-\dot{x}_0{g_2\over B}\sin\phi )
$$
$$
+ {1\over B}(-\dot{x}_0g_3\cos\phi -\dot{y}_0g_3\sin\phi)~~.
$$
Collecting terms of the form
$$
{1\over 2\pi}\int_0^{2\pi}\cos^n\phi\sin^m\phi d\phi,~~~m,n~~\mbox{even}
$$
we find that $<f_2>=0$, while
\begin{eqnarray}\label{f3anal}
<f_3> &= &{1\over 16b^2c\sin^2(1+c)t_0}
\Bigg(4abc^{1/2}\sin ct(-{\dot{x}_0\over x_0^2}+b^2c\cos(1+c)t\dot{y}_0) \nonumber\\
& &+4\dot{x}_0\dot{y}_0(b^4c^2x_0^4-{1\over x_0^4})
-4bc^{1/2}\cos(1+c)t(\dot{x}_0^2-\dot{y}_0^2)({1\over x_0^2}+b^2cx_0^2)   \Bigg)\nonumber\\
& &+{bc^{1/2}\over 4bc^{1/2}\sin(1+c)t_0}
\Bigg({\dot{y}_0\cos(1+c)t_0\over x_0}-bc^{1/2}x_0\dot{x}_0   \Bigg)
\end{eqnarray}
which after some simplification yields Eq.(\ref{f3}).
\end{document}